\documentclass[prb,reprint, amsmath,amssymb,showpacs,superscriptaddress,floatfix]{revtex4-2}
\usepackage[breaklinks=true,colorlinks,citecolor=blue,linkcolor=blue,urlcolor=blue]{hyperref}
\usepackage[table,xcdraw]{xcolor}
\usepackage{epsfig,mathrsfs,color,latexsym,subfigure,marginnote,gensymb}
\usepackage{graphicx}
\usepackage{xcolor}
\usepackage{booktabs}
\usepackage{appendix}

\begin{document}
\title{Terahertz Magnon Excitations and Switching in Non-Collinear Antiferromagnets}
\author{Durga Prasad Goli}
\email{durga@kaist.ac.kr}
\affiliation{Department of Physics, Korea Advanced Institute of Science and Technology, Daejeon 34141, Republic of Korea}
\author{Se Kwon Kim}
\email{sekwonkim@kaist.ac.kr}
\affiliation{Department of Physics, Korea Advanced Institute of Science and Technology, Daejeon 34141, Republic of Korea}
\date{\today}

\begin{abstract}

We investigate how spatiotemporal spin polarized current can lead to terahertz frequency excitations in non-collinear antiferromagnets. By solving the Landau-Lifshitz-Gilbert equation numerically for non-collinear antiferromagnet, we show that the magnon frequency spectrum exhibits standing spin wave modes and depends on the thickness of Mn$_3$Ge in heterostructure Fe\textbar Au\textbar Mn$_3$Ge. Also, we analyze the switching process of ground state as a function of a spin current. We show a switching phase diagram, which contains switching and non-switching regions. Our work suggests non-collinear antiferromagnets as an efficient platform for terahertz magnonics and ultrafast memory devices.

\end{abstract}

\maketitle

\section{Introduction}

The discovery of magnetization quenching \cite{Beaurepaire1996}, excitation of terahertz (THz) frequency waves \cite{Kirilyuk2013,Kampfrath2013,Seifert2016}, and switching \cite{Stanciu2007} through femtosecond laser pulses paved the way for the emergence of femtomagnetism as a new field \cite{Walowski2016,Malinowski2018}.
The potential of femtosecond laser pulses to provide further insights into ultrafast magnetization dynamics has generated considerable interest. They offer new opportunities for controlling magnetization in thin magnetic layers within picoseconds.
Application of a femtosecond laser pulse on a ferromagnet\textbar nonmagnet (FM\textbar NM) structure causes majority spin electrons to exit FM, resulting in demagnetization within a few picoseconds \cite{Eschenlohr2013,Bergeard2016}. The superdiffusion electron transport theory, introduced by Battiato et al. \cite{Battiato2010,Battiato2012,Balaz2018}, explained this phenomenon. Similarly, in a spin valve comprising of two ferromagnets (FM1 and FM2) and a nonmagnetic spacer (FM1\textbar NM\textbar FM2), the polarized spin carriers can reach FM2 due to demagnetization in FM1 by femtosecond laser pulse \cite{Razdolski2017,Ulrichs2018,Ritzmann2020}. The ultrafast polarized spin current exerts spin transfer torque (STT) on FM2, resulting the generation of THz dynamics and switching magnetization. Recent studies on ferromagnet heterostructures revealed that the excitation of magnon frequencies reportedly reached up to a few THz \cite{Razdolski2017,Ulrichs2018,Ritzmann2020}.

To enhance the excitation of magnon frequency range, antiferromagnets (AFM) are considered to be reliable choice over FMs. They offer several advantages, including minimal stray fields, ultrafast magnetic dynamics, and higher magnon frequencies \cite{Jungwirth2016,Baltz2018}. Ultrafast THz dynamics of the N{\'e}el vector in AFMs can be induced by spin currents \cite{Cheng2014} and light \cite{Kimel2009,Kampfrath2011,Hortensius2021,Ross2024}. Theoretical predictions on collinear AFM revealed that ultrafast spin dynamics caused by a spin current from a femtosecond laser pulse can excite resonant magnon modes with frequencies higher than FMs \cite{Chirac2020,Weibenhofer2023}.

Going beyond conventional collinear AFMs, non-collinear antiferromagnetic (NAFM) materials like Mn$_3$X (X=Ir, Sn, Ge) are capturing interest due to their unconventional transport phenomena such as the anomalous Hall effect \cite{Chen2014,Kubler_2014,Nakatsuji2015,Ajaya2016}, anomalous Nernst effect \cite{Ikhlas2017,Reichlova2019}, and magneto-optical Kerr effect \cite{Higo2018}.
Recent studies revealed that the ultrafast dynamics \cite{Shukla2022} and deterministic switching \cite{Tsai2020} within the influence of STT in Mn$_3$Sn. In this paper, we demonstrate the generation of standing spin waves with high-frequency magnon modes capable of reaching several THz by considering Mn$_3$Ge. Taking advantage of the ultrafast N{\'e}el vector dynamics in Mn$_3$Ge, the switching mechanism facilitates reversal of N{\'e}el vector within  a few picoseconds.

\begin{figure}
\centering
\includegraphics[width=0.49\textwidth]{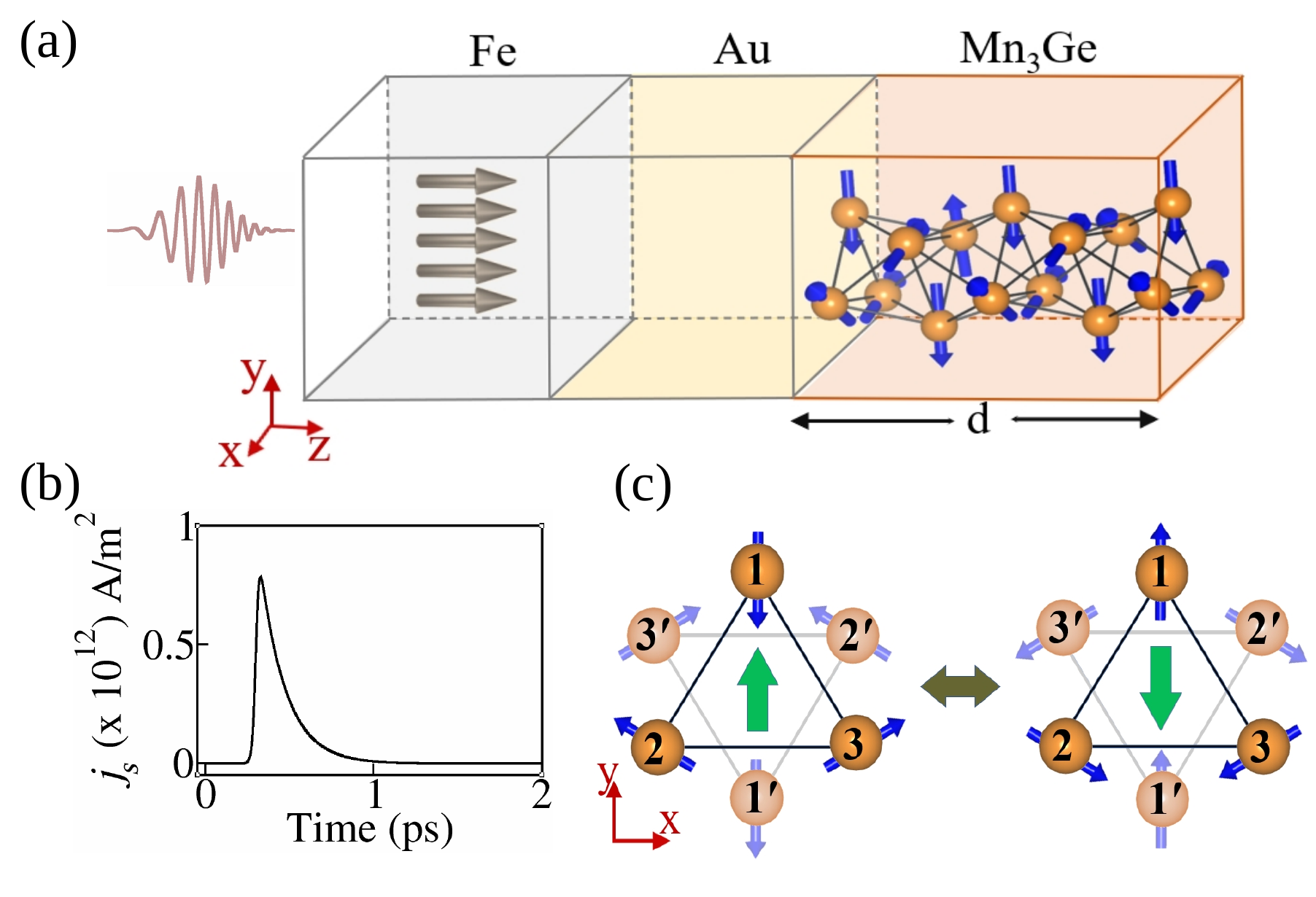}
\caption{(a) Schematic illustration of Fe\textbar Au\textbar Mn$_3$Ge spin valve heterostructure, where a femtosecond laser pulse irradiates Fe. (b) The resultant temporal polarized spin current at Mn$_3$Ge using Eq. (\ref{eq_js}) with $j_0$=10$^{12}$ A/m$^2$. (c) Two degenerate ground states of Mn$_3$Ge, each with a magnetic unit cell of six Mn atoms within kagome bilayer and the corresponding spin texture is a 120$^\circ$ non-collinear ground state. Collective magnetic order of the ground state is represented by octupole moment (green arrow). It is possible to switch from one ground state to another using $j_s$.} 
\label{fig_1}
\end{figure}
The structure of this paper is as follows. In Sec. II, we introduce a spin valve heterostructure along with the spin Hamiltonian. We describe atomistic spin simulation technique for computing the ultrafast spin dynamics due to STT on Mn$_3$Ge. In Sec. III, we investigate magnon spectrum and switching phase diagrams. We show that the emergence of standing spin waves, whose resonance frequencies can reach THz range as a function of thickness. We demonstrate that the STT induces ultrafast dynamics in the N{\'e}el vector, enabling the field-free switching in Mn$_3$Ge. In Sec. IV, we summarize the results.

\section{Model and Spin Hamiltonian}

We propose a heterostructure spin valve, comprising Fe\textbar Au\textbar Mn$_3$Ge as shown in Fig. \ref{fig_1}(a).
In this structure, we consider a thin film of Mn$_3$Ge, which is made up of AB stacked kagome lattice planes in the [0001] direction \cite{Kiyohara2016}. The presence of geometric frustration causes a ground state to be NAFM order with sublattice spins set at 120$^\circ$. 
Fig. \ref{fig_1}(c) shows that a magnetic unit cell of six Mn atoms within a kagome bilayer. The sublattice spins in layer-1 are labeled as 1,2 and 3 whereas 1$^\prime$, 2$^\prime$ and 3$^\prime$ for layer-2. The corresponding spin texture is characterized by a ferroic order of cluster magnetic octupole \cite{Nomoto2020,Dasgupta2022}.
The spin Hamiltonian describing Mn$_3$Ge is given by
\begin{align}
        \mathcal{H}~=~&\frac{J_{\textrm{AFM}}^{\textrm{intra}}}{2}\sum_{\langle i,j \rangle} \vec{S}_i \cdot \vec{S}_j
	~+~\frac{J_{\textrm{FM}}^{\textrm{inter}}}{2}\sum_{\langle i,j^\prime \rangle} \vec{S}_i \cdot \vec{S}_{j^\prime} \nonumber \\
	&+~\frac{J_{\textrm{AFM}}^{\textrm{inter}}}{2}\sum_{\langle i,j^\prime \rangle} \vec{S}_i \cdot \vec{S}_{j^\prime}
        ~+~\frac{D^{\textrm{intra}}}{2}~\sum_{\langle i,j \rangle}{\hat z}\cdot(\vec{S}_i \times \vec{S}_j) \nonumber \\
        &-~K\sum_i(\vec{S}_i \cdot {\hat n}_i)^2,
\label{eq_ham}
\end{align}\\
\noindent where the local magnetic moment of Mn at site $i$ is defined by unit spin vector ${\vec S}_i$ with magnitude $\mu_s$.
$\langle i,j \rangle$ and $\langle i,j^\prime \rangle$ represent bonding between intralayer and interlayer spins.
$J_{\textrm{AFM}}^{\textrm{intra}}$ is the intralayer nearest-neighbor AFM exchange interaction strength.
The interlayer FM and AFM exchange strengths are represented by $J_{\textrm{FM}}^{\textrm{inter}}$ and $J_{\textrm{AFM}}^{\textrm{inter}}$ respectively. The intralayer Dzyaloshinskii-Moriya interaction is $D^{\textrm{intra}}$ and $K$ is the easy-axis in-plane anisotropy, whose sublattice vectors are ${\hat n}_1=-{\hat y}$, ${\hat n}_2=-({\sqrt 3}/2){\hat x}+(1/2){\hat y}$ and ${\hat n}_3=({\sqrt 3}/2){\hat x}+(1/2){\hat y}$ (shown in Fig. \ref{fig_1}(c)). The corresponding spin configuration is one of the degenerate ground states of Mn$_3$Ge. For simulations, the parameters in Eq. (\ref{eq_ham}) are taken as : $J_{\textrm{AFM}}^{\textrm{intra}}$ = 37.4 meV, $J_{\textrm{FM}}^{\textrm{inter}}$ = -18.7 meV, $J_{\textrm{AFM}}^{\textrm{inter}}$ = 6.24 $\times$ 10$^{-3}$ meV, $D^{\textrm{intra}}$ = 2.18$\times$10$^{-2}$ meV and $K$ = 10$^{-3}$ meV \cite{Chen2020,Chaudhary2022}. The interlayer distance between two kagome planes is taken as 2.2 \textup{~\AA}.
In this work, the size of Mn$_3$Ge system is taken to be 4$\times$4$\times$$d$ nm$^3$, where $d$ is thickness of Mn$_3$Ge. Periodic boundary conditions are implemented in the xy-plane.

The spin valve setup shown in Fig. \ref{fig_1}(a) utilizes the superdiffusive transport theory \cite{Battiato2010,Battiato2012}, implies Fe absorbs a femtosecond laser pulse, resulting in the excitation of spin electrons, polarized in the magnetization direction of Fe. To enhance the effect of STT due to polarized spin current, we assume the easy-plane of Mn$_3$Ge is perpendicular to the magnetization direction of Fe. 
Utilizing the Landau-Lifshitz-Gilbert (LLG) equation, we carry out atomistic spin simulations on Mn$_3$Ge in the presence of STT. The LLG equation is integrated numerically with the support of fourth order Runge-Kutta solver. The LLG equation is expressed as follows:
\begin{align}
\alpha {\vec S}_i \times \frac{\partial{\vec S}_i}{\partial t
	}~-~\frac{\partial{\vec S}_i}{\partial t}~=~&\gamma {\vec S}_i \times {\vec H}_i^{\textrm{eff}} \nonumber \\
	&+~\frac{\gamma \hbar \theta}{2edM_s}j_s [{\vec S}_i \times ({\vec S}_i \times {\hat z})],
\label{eq_llg}
\end{align} \\
\noindent where $\gamma$ is the gyromagnetic ratio and $\alpha$ is the intrinsic Gilbert damping. The effective field of site $i$ is represented by ${\vec H}_i^{\textrm{eff}}=-\frac{1}{\mu_s}\frac{\partial \mathcal{H}}{\partial {\vec S}_i}$. Other parameters are as follows: $\theta$ is the spin Hall angle, $e$ is charge of an electron, $d$ is thickness, $M_{s}$ is saturation of magnetization and ${\hat z}$ is the polarization direction of spin current. 
$j_s$ is the spin current due to femtosecond laser pulse absorbtion of Fe. In this model, $j_s$ is considered to be spatially dependent in Mn$_3$Ge as it transfers spin angular momentum to local Mn spins, impacting its magnitude \cite{Razdolski2017}. Thus, the profile of $j_s$ is presented in a spatiotemporal form as shown below:
\begin{align}
j_s~=~j_0 e^{-z/\lambda_{\textrm{STT}}} \frac{e^{-t/\tau_2}}{1+e^{-(t-t_0)/\tau_1}}.
\label{eq_js}
\end{align} \\
\noindent Here $j_0$ is the magnitude of spin current at Au\textbar Mn$_3$Ge interface, $z$ is length of Mn$_3$Ge and $\lambda_{\textrm{STT}}$ is penetration depth of spin current. The parameters $t_0$, $\tau_1$ and $\tau_2$ characterize temporal profile of spin current.
$t_0$ represents delay time of pulse, while $\tau_1$ and $\tau_2$ are exponential decay times.
The values given to the parameters in Eqs. (\ref{eq_llg}) and (\ref{eq_js}) are as follows: $\alpha$ = 0.003, $M_{s}$ = 2.23$\times$10$^5$ A/m \cite{Shukla2022} and $\mu_s$=2.5 $\mu_B$. The spin Hall angle is set to be 1, adopted from Refs. \cite{Alekhin2017,Ulrichs2018} and $\lambda_{\textrm{STT}}$=1 nm to mimic experimental conditions \cite{Ghosh2012}.
The time step for numerical integration of Eq. (\ref{eq_llg}) is 1 fs and total integration time is 500 ps. All the calculations are done at T = 0 K. The temporal variation of spin current is shown in Fig. \ref{fig_1}(b) for $j_0$ = 10$^{12}$ A/m$^2$ and temporal parameters $t_0$ = 300 fs, $\tau_1$ = 10 fs and $\tau_2$ = 150 fs for first layer of Mn$_3$Ge.

\begin{figure}
\centering
\includegraphics[width=0.49\textwidth]{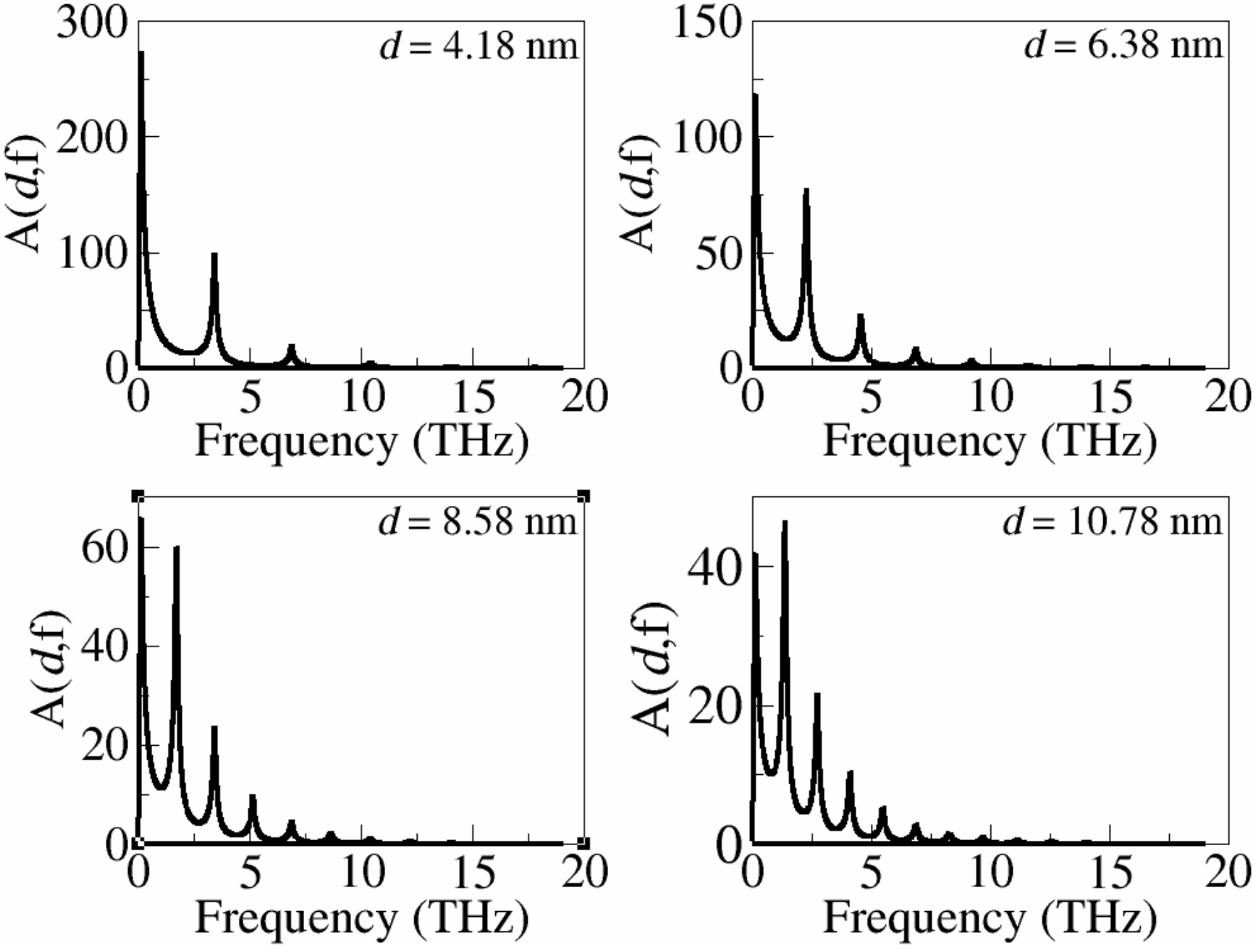}
	\caption{Excited magnon spectrum $A(z,f)$ [Eq. (\ref{eq_fft})] at the most distant layer ($z$ = $d$) from the spin-current source for four different thicknesses, $d$ = 4.18, 6.38, 8.58, 10.78 nm. The number of magnon modes increases with the thickness.}
\label{fig_2}
\end{figure}

\section{Results}

First, using the atomistic spin simulation technique, we relax the spins of Mn$_3$Ge to the ground state as described in Eq. (\ref{eq_ham}) without spin current. 
This can be done by integrating Eq. (\ref{eq_llg}) numerically with $j_0$ = 0. After initial equilibration, we turn on $j_s$ by taking $j_0$ = 10$^{12}$ A/m$^2$ to investigate the spin current effect on Mn$_3$Ge as a function of thickness. We set $t_0$ = 300 fs, $\tau_1$ = 10 fs and $\tau_2$ = 150 fs. STT process excites ultrafast dynamics of all Mn spins in Mn$_3$Ge, leading to the generation of high-frequency spin waves that propagate throughout the sample. These spin waves reflect multiple times at the open ends over hundreds of picoseconds, resulting in the formation of standing spin waves at THz frequencies.
The Mn spins exhibit in-plane precession, while maintaining a non-zero z-component of the spin vector. The spin precessions include both degenerate and highly excited spin configurations. 
By employing the atomistic spin simulations, we examine the time evolution of z-component of each spin ($S_i^z(z,t)$) (irrespective of the sublattice and layer) for 500 ps time interval. The Fast Fourier transformations (FFT) are carried out for $S_i^z(z,t)$ to extract the frequency spectrum. 
The amplitudes of FFT  are calculated by taking the $S_i^z$ of each layer is given by
\begin{align}
	A(z,f)~=~\sum_{i=1}^{M} \sum_{j=1}^{N_{\textrm{steps}}}S_i^z(z,t)e^{-i2\pi ft},
\label{eq_fft}
\end{align} \\
\noindent where $S_i^z(z,t)$ is z-component of $i^{th}$ site spin as a function of thickness coordinate ($z$) and time $t$. 
$N_{\textrm{steps}}$ is the number of integration time steps and $M$ is number of spins per layer.

\begin{figure}
\centering
\includegraphics[width=0.48\textwidth]{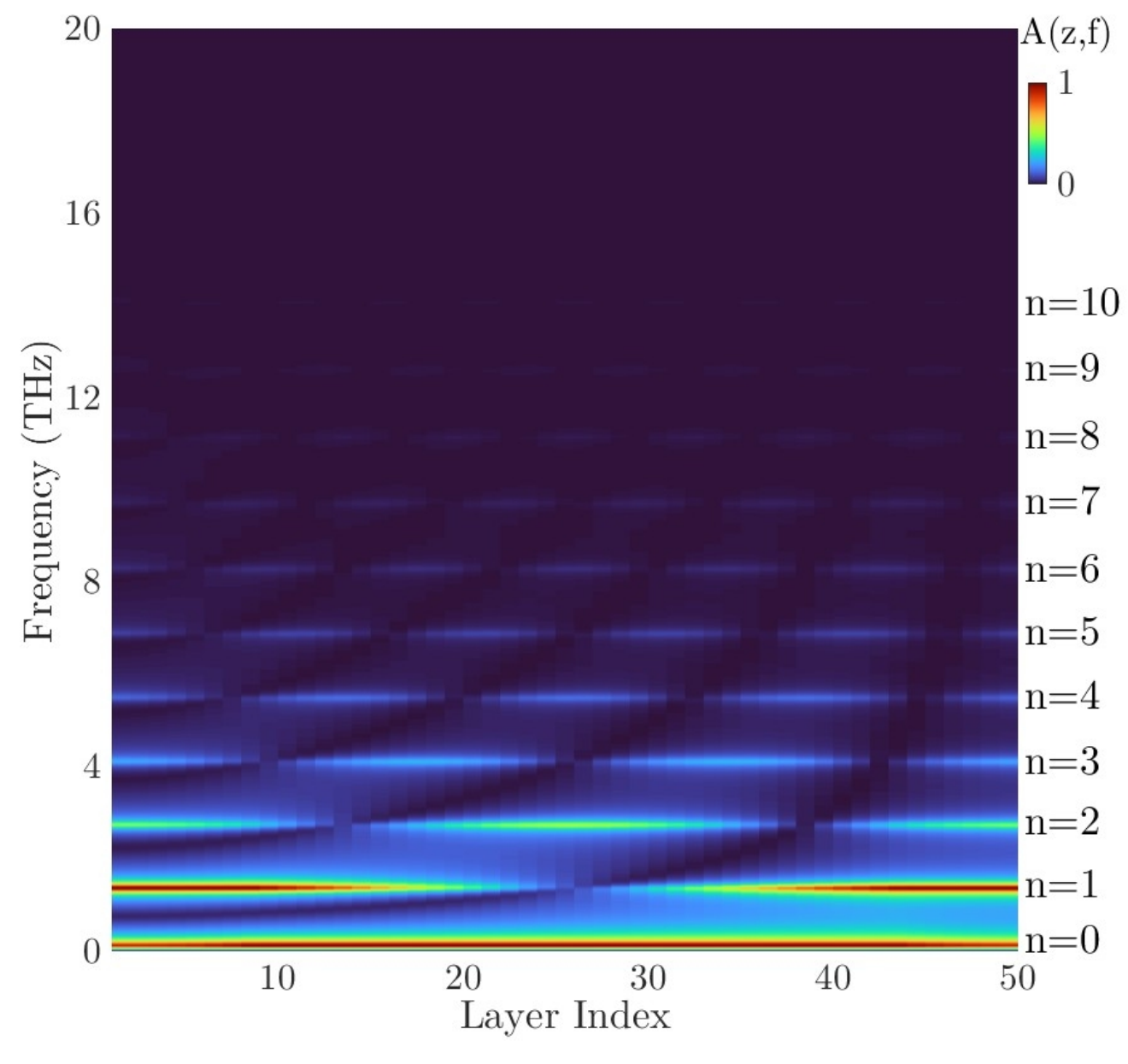}
\caption{Layer dependent magnon spectrum $A(z,f)$ [Eq. (\ref{eq_fft})] for the thickness, d = 10.78 nm with 50 layers. The number of standing spin wave modes, $n$ = 0,1,2,...,10 are appeared. Here A(z,f) is the normalized FFT amplitude.}
\label{fig_3}
\end{figure}

Spin excitations triggered by $j_s$ lead to maximum spin dynamics at the resonance frequencies of Mn$_3$Ge, causing frequency peaks to emerge with large amplitude compared to the rest of the spectrum.  To investigate the spin wave excitations, we calculate the magnon spectrum of spins in the last layer of Mn$_3$Ge for different thicknesses, shown in Fig. \ref{fig_2}. 
Distinct peaks observed in each spectrum shows the formation of standing spin waves at resonant frequencies, regardless of film thickness. For a film of finite thickness, these standing spin waves are represented by a wave vector, $k_n=n\pi/d$ ($n$ being an integer). 
The magnon spectrum span a broad range of THz frequencies and are resonantly excited with wave vectors satisfy the confinement condition of $k_n$.
The low frequency peak ($n=0$) is the fundamental AFM resonance frequency of Mn$_3$Ge, which is 0.13 THz from Fig. \ref{fig_2}. This peak position remains constant regardless of the thickness. As $d$ increases, more peaks start to appear in the magnon spectrum and the separation between peaks decreases. Also, the peaks with $n~\textgreater~0$ move towards lower frequencies. The amplitudes of higher frequency peaks are noticeably smaller compared to the peaks in the low frequency spectrum. This is due to fact that the spin dynamics of high frequencies being limited to a few picoseconds, while lower frequency dynamics can last for a few hundred picoseconds. 

We now examine the spatial distribution of spin-wave modes in the Mn$_3$Ge. Fig. \ref{fig_3} illustrates the spatial dependent magnon spectrum for $d$ = 10.78 nm with 50 layers. 
The FFTs are carried out for $S_i^z(z,t)$ for each layer separately over a duration of 500ps to get the layer dependent spectrum.
The emergence of standing spin wave spectrum in Mn$_3$Ge is a result of constructive interference among the reflected spin waves. Apart from the fundamental AFM resonance mode, there are 10 visible spin wave nodes ($n$ = 10) with  wavelength, $\lambda =2d/n$. The magnon spectrum does not show higher frequency modes ($n~\textgreater~10$). This is due to the fact that magnon excitations of these modes have short lifetimes and are suppressed quickly by damping \cite{Razdolski2017}. The standing spin wave magnon spectrum resembles the spectrum of FM and collinear AFM \cite{Ulrichs2018,Ritzmann2020,Weibenhofer2023}. 

\begin{figure}
\centering
\includegraphics[width=0.48\textwidth,height=0.45\textwidth]{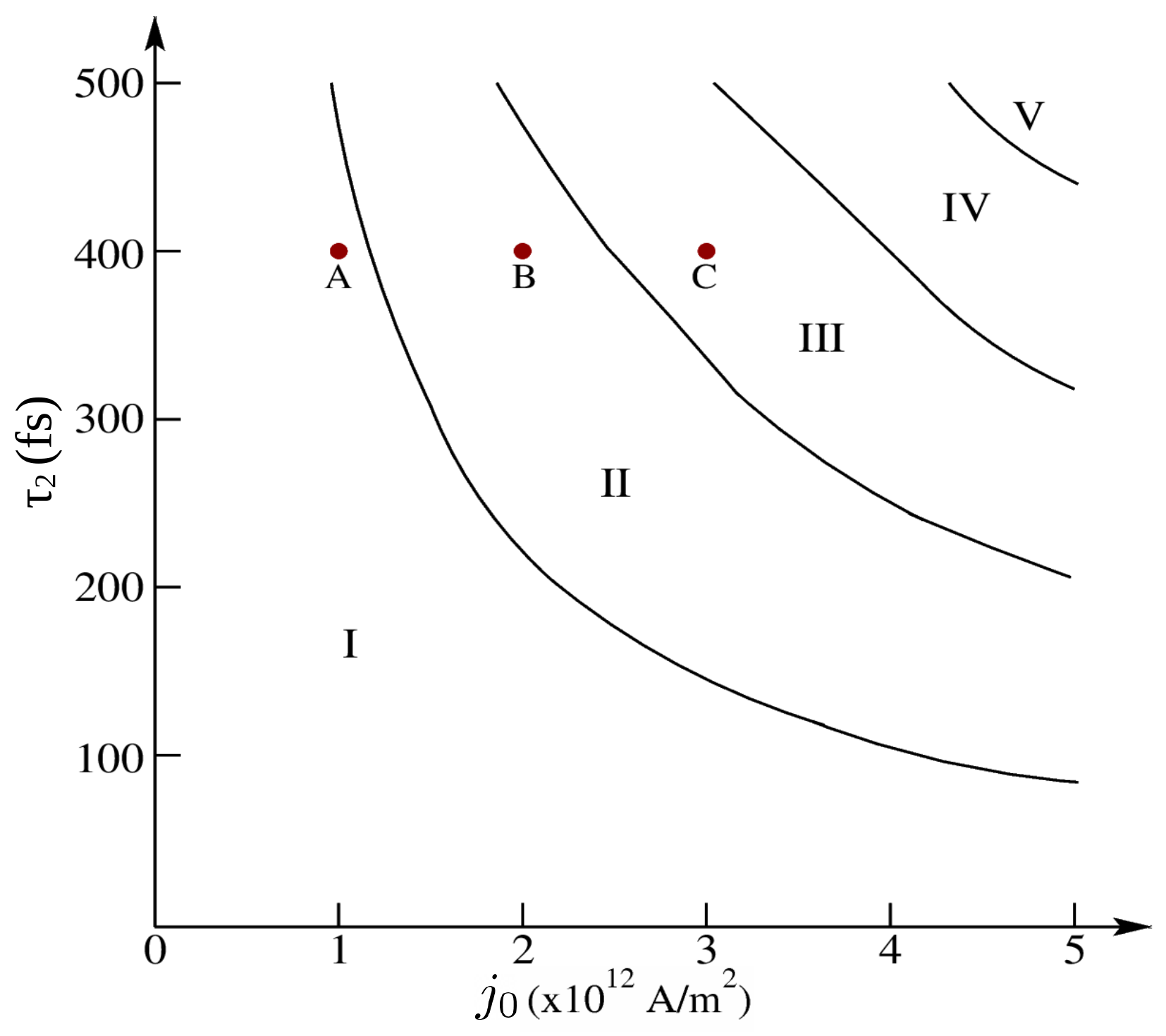}
\caption{Switching phase diagram as a function of $j_0$ and $\tau_2$. This plot consists of switching (II and IV) and non-switching (I, III and V) regions.}
\label{fig_4}
\end{figure}
Now we present the results for switching process in Mn$_3$Ge. 
This can be identified by observing the sign change of N{\'e}el vector and change in the octupole moment direction. For a three sublattice system, the two N{\'e}el vectors can be defined as follows: ${\vec l}_1=({\vec S}_1 + {\vec S}_2 - 2{\vec S}_3)/3{\sqrt{2}}$ and ${\vec l}_2=(-{\vec S}_1 + {\vec S}_2)/{\sqrt{6}}$ \cite{Gomonay2015,Goli2021}. Below, we show our simulation results in terms of one N{\'e}el vector (${\vec l}_1$) and the angle of the octupole moment with y-axis, which is denoted by $\phi_{\textrm{oct}}$. Since it is observed that, during the spin dynamics driven by STT, the spins constituting the system preserve the 120$^\circ$ non-collinear configuration, we take the mirror image of $\mathbf{S}_1$ with respect to the $xz$-plane as the octupole moment \cite{He2024}. As the spins rotate clockwise, the octupole moment rotates counter clockwise, fulfilling the condition of opposite handedness \cite{Yoon2023}.
To observe a successful switching process, spins should gain enough energy due to $j_s$ to overcome the anisotropy energy barrier into a new spin reversal configuration \cite{Chaudhary2022}.

\begin{figure}
\centering
\includegraphics[width=0.49\textwidth,height=0.405\textwidth]{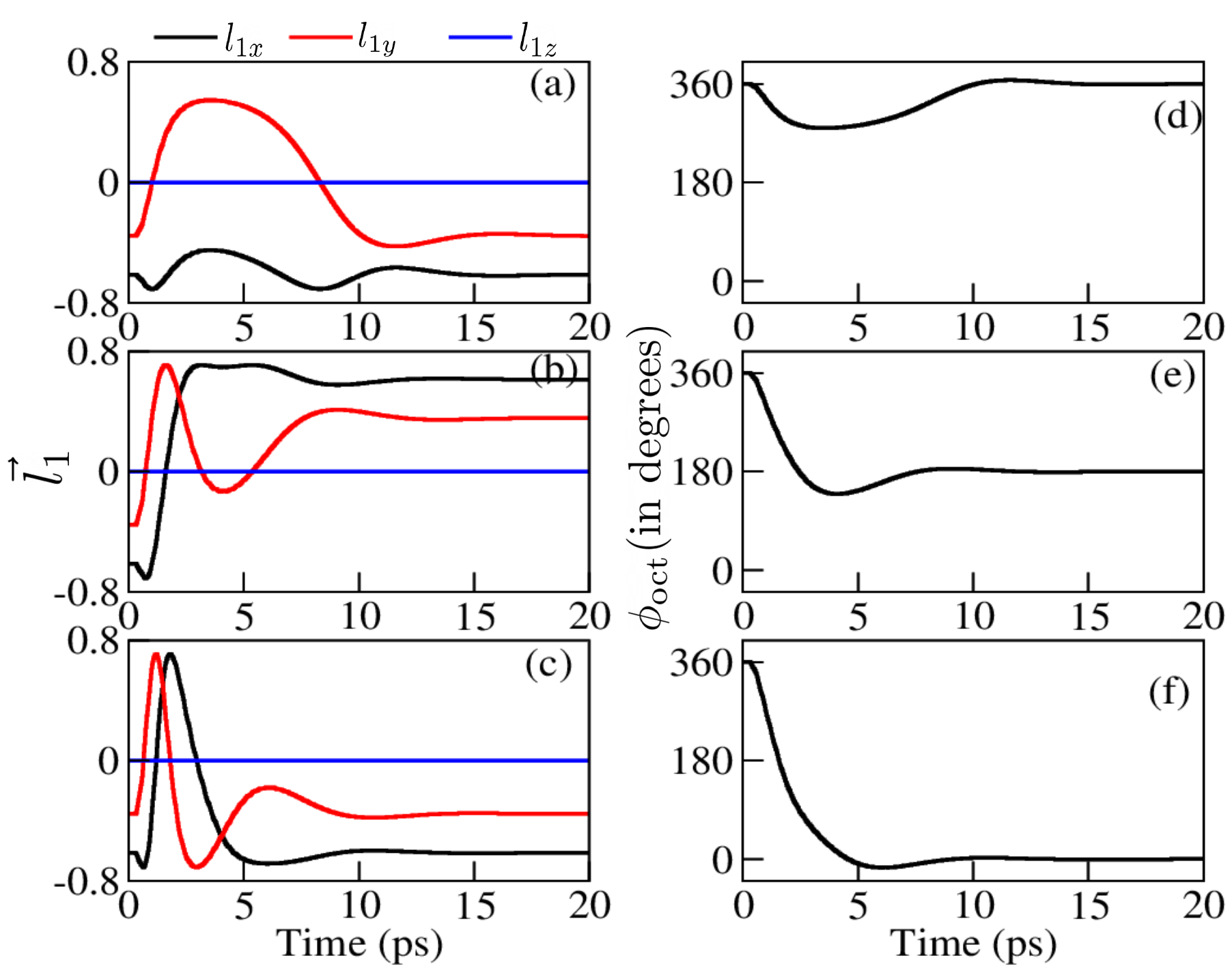}
	\caption{The dynamics of the N{\'e}el vector, ${\vec l}_1$ and the angle of the magnetic octupole moment, ${\phi}_{\textrm{oct}}$  (with respect to the y-axis) are shown for three different $j_0$ values, taken from Fig. \ref{fig_4} indicated in red dots and are labeled as A, B and C respectively. Here $j_0$ = 1$\times$10$^{12}$ A/m$^2$ for (a),(d), $j_0$ = 2$\times$10$^{12}$ A/m$^2$ for (b),(e) and $j_0$ = 3$\times$10$^{12}$ A/m$^2$ for (c),(f). Successful switching is observed for $j_0$ = 2$\times$10$^{12}$ A/m$^2$ and is shown in (b) and (e). However, no switching process is observed in (a),(d) and (c),(f). The values of ${\phi}_{\textrm{oct}}$ below 0$^\circ$ and above 360$^\circ$ can be translated as ${\phi}_{\textrm{oct}}$ = ${\phi}_{\textrm{oct}}$ + 360$^\circ$ and ${\phi}_{\textrm{oct}}$ = ${\phi}_{\textrm{oct}}$ - 360$^\circ$ respectively.} 
\label{fig_5}
\end{figure}

For the switching calculations, we keep $t_0$ = 300 fs and $\tau_1$ = 10 fs fixed while varying $j_0$ and $\tau_2$.
Fig. \ref{fig_4} illustrates the switching phase diagram as a function of $j_0$ and $\tau_2$ for $d$ = 4.18 nm. 
This phase diagram consists of switching and non-switching regions. From the figure, it is evident that critical values of $j_0$ and $\tau_2$ are essential for the switching process. For a better understanding, we calculate the time evolution of ${\vec l}_1$ and $\phi_{\textrm{oct}}$ for a few points from Fig. \ref{fig_4}, which are labeled by A, B and C with different $j_0$ and $\tau_2$ = 400 fs.
No switching is observed at A in region I for $j_0$ = 1$\times$10$^{12}$ A/m$^2$ and magnetic unit cell returns to the initial state. 
Only subtle dynamics observed in ${\vec l}_1$ and $\phi_{\textrm{oct}}$, recovering their initial values within a few picoseconds as shown in Figs. \ref{fig_5}(a) and (d).
This is due to the temporal profile of $j_s$ is not sufficient enough to exert STT on spins to overcome the barrier. 

To realize a successful switching process, sufficiently large values of $j_0$ and $\tau_2$ are required to overcome the barrier between the degenerate ground states. 
At point B in region-II, a successful switching is observed for $j_0$ = 2$\times$10$^{12}$ A/m$^2$, with all spins rotating 180$^\circ$ from the initial ground state. As shown in Figs. \ref{fig_5}(b) and (e), sign of ${\vec l}_1$ is reversed and a 180$^\circ$ rotation of octupole moment leads to a 180$^\circ$ change in $\phi_{\textrm{oct}}$.
The initial and final spin configurations of a magnetic unit cell are shown in Fig. \ref{fig_1}(c).
In regions II and IV, successful switching process occurs place for all ($j_0$, $\tau_2$). In these regions, switching time depends on the magnitude of $j_0$ as large $j_0$ potentially causes shorter switching time. This illustrates one of our main results: non-collinear magnets can be switched with suitable choice of the magnitude and the pulse duration of the spin current. In region III at point C with $j_0$ = 3$\times$10$^{12}$ A/m$^2$, spins return to their initial state via a full 360$^\circ$ rotation, resulting in no switching. Figs. 5(c) and (f) show that  ${\vec l}_1$ and $\phi_{\textrm{oct}}$ return to their initial values. The spin dynamics videos of magnetic unit cell at A, B and C are provided in Supplemental Material.
Similar to region-III, spins return to their initial states in region V indicates that no switching process occurs in these regions.
To ensure a successful switching process with increasing $d$, larger values of $j_s$ are needed.
It is also possible to observe double switching by employing two $j_s$ pulses with a time delay of a few picoseconds between the pulses for a the combination of ($j_0$, $\tau_2$) in region II and IV. 
In this process, the initial spin configuration is reversed by the first pulse; a second pulse rotates all spins by another 180$^\circ$, returning the system to its initial state.
Recent studies revealed that the spin reversal switching other than 180$^\circ$ is possible in the presence external field \cite{Tsai2020,Xu2024}.

\section{Discussion}

The main objective of this research was theoretical investigation of THz magnon excitations in NAFM. 
Our study predicted that it is possible to observe THz frequency excitations in Mn$_3$Ge. Similar behavior was observed in Fe and Mn$_2$Au \cite{Ulrichs2018,Ritzmann2020,Weibenhofer2023} and the resonant frequency spectrum spans a few THz. Despite that, our studies revealed the use of NAFMs could be advantageous for boosting the resonant frequencies higher than 10 THz. Manipulating the film thicknesses can lead to the control and excitation of resonant frequencies above the fundamental mode. We suggest that thickness can be as a tool to control resonant THz frequencies of magnetic materials.

We also showed the switching mechanism in Mn$_3$Ge using spin current due to femtosecond laser pulse. It is possible to control the switching process by temporal profile of femtosecond laser pulse. 
An investigation was conducted on how damping affects the switching time of the N{\'e}el vector in Mn$_2$Au \cite{Weibenhofer2023}. In the case of Mn$_3$Ge, two damping values were considered for spin dynamics \cite{Shukla2022,Chaudhary2022}. Therefore, it is essential to continue the investigation of switching process until experimental observations are revealed. 

In conclusion, we have showed the emergence of standing spin waves through the magnon excitations in NAFM with spatiotemporal polarized spin current pulse. The magnon spectrum revealed that the excitations of THz frequency spin waves along with fundamental AFM resonance frequency. 
The thickness of Mn$_3$Ge determines how many frequency modes are present in the spin wave spectrum. Furthermore, we illustrated the switching of ground state within a few picoseconds timescale. Our work may be helpful towards THz magnonics and picosecond switching of NAFM based spintronic devices.

\begin{acknowledgments}
This work was supported by Brain Pool Plus Program through the National Research Foundation of Korea funded by the Ministry of Science and ICT (NRF-2020H1D3A2A03099291).
\end{acknowledgments}

\bibliography{ref}

\begin{thebibliography}{45}%
\makeatletter
\providecommand \@ifxundefined [1]{%
 \@ifx{#1\undefined}
}%
\providecommand \@ifnum [1]{%
 \ifnum #1\expandafter \@firstoftwo
 \else \expandafter \@secondoftwo
 \fi
}%
\providecommand \@ifx [1]{%
 \ifx #1\expandafter \@firstoftwo
 \else \expandafter \@secondoftwo
 \fi
}%
\providecommand \natexlab [1]{#1}%
\providecommand \enquote  [1]{``#1''}%
\providecommand \bibnamefont  [1]{#1}%
\providecommand \bibfnamefont [1]{#1}%
\providecommand \citenamefont [1]{#1}%
\providecommand \href@noop [0]{\@secondoftwo}%
\providecommand \href [0]{\begingroup \@sanitize@url \@href}%
\providecommand \@href[1]{\@@startlink{#1}\@@href}%
\providecommand \@@href[1]{\endgroup#1\@@endlink}%
\providecommand \@sanitize@url [0]{\catcode `\\12\catcode `\$12\catcode
  `\&12\catcode `\#12\catcode `\^12\catcode `\_12\catcode `\%12\relax}%
\providecommand \@@startlink[1]{}%
\providecommand \@@endlink[0]{}%
\providecommand \url  [0]{\begingroup\@sanitize@url \@url }%
\providecommand \@url [1]{\endgroup\@href {#1}{\urlprefix }}%
\providecommand \urlprefix  [0]{URL }%
\providecommand \Eprint [0]{\href }%
\providecommand \doibase [0]{https://doi.org/}%
\providecommand \selectlanguage [0]{\@gobble}%
\providecommand \bibinfo  [0]{\@secondoftwo}%
\providecommand \bibfield  [0]{\@secondoftwo}%
\providecommand \translation [1]{[#1]}%
\providecommand \BibitemOpen [0]{}%
\providecommand \bibitemStop [0]{}%
\providecommand \bibitemNoStop [0]{.\EOS\space}%
\providecommand \EOS [0]{\spacefactor3000\relax}%
\providecommand \BibitemShut  [1]{\csname bibitem#1\endcsname}%
\let\auto@bib@innerbib\@empty
\bibitem [{\citenamefont {Beaurepaire}\ \emph {et~al.}(1996)\citenamefont
  {Beaurepaire}, \citenamefont {Merle}, \citenamefont {Daunois},\ and\
  \citenamefont {Bigot}}]{Beaurepaire1996}%
  \BibitemOpen
  \bibfield  {author} {\bibinfo {author} {\bibfnamefont {E.}~\bibnamefont
  {Beaurepaire}}, \bibinfo {author} {\bibfnamefont {J.-C.}\ \bibnamefont
  {Merle}}, \bibinfo {author} {\bibfnamefont {A.}~\bibnamefont {Daunois}},\
  and\ \bibinfo {author} {\bibfnamefont {J.-Y.}\ \bibnamefont {Bigot}},\
  }\bibfield  {title} {\bibinfo {title} {Ultrafast spin dynamics in
  ferromagnetic nickel},\ }\href {https://doi.org/10.1103/PhysRevLett.76.4250}
  {\bibfield  {journal} {\bibinfo  {journal} {Phys. Rev. Lett.}\ }\textbf
  {\bibinfo {volume} {76}},\ \bibinfo {pages} {4250} (\bibinfo {year}
  {1996})}\BibitemShut {NoStop}%
\bibitem [{\citenamefont {Kirilyuk}\ \emph {et~al.}(2013)\citenamefont
  {Kirilyuk}, \citenamefont {Kimel},\ and\ \citenamefont
  {Rasing}}]{Kirilyuk2013}%
  \BibitemOpen
  \bibfield  {author} {\bibinfo {author} {\bibfnamefont {A.}~\bibnamefont
  {Kirilyuk}}, \bibinfo {author} {\bibfnamefont {A.~V.}\ \bibnamefont
  {Kimel}},\ and\ \bibinfo {author} {\bibfnamefont {T.}~\bibnamefont
  {Rasing}},\ }\bibfield  {title} {\bibinfo {title} {Laser-induced
  magnetization dynamics and reversal in ferrimagnetic alloys},\ }\href
  {https://doi.org/10.1088/0034-4885/76/2/026501} {\bibfield  {journal}
  {\bibinfo  {journal} {Rep. Prog. Phys.}\ }\textbf {\bibinfo {volume} {76}},\
  \bibinfo {pages} {026501} (\bibinfo {year} {2013})}\BibitemShut {NoStop}%
\bibitem [{\citenamefont {Kampfrath}\ \emph {et~al.}(2013)\citenamefont
  {Kampfrath}, \citenamefont {Battiato}, \citenamefont {Maldonado},
  \citenamefont {Eilers}, \citenamefont {N\"otzold}, \citenamefont
  {M\"ahrlein}, \citenamefont {Zbarsky}, \citenamefont {Freimuth},
  \citenamefont {Mokrousov}, \citenamefont {Bl\"ugel}, \citenamefont {Wolf},
  \citenamefont {Radu}, \citenamefont {Oppeneer},\ and\ \citenamefont
  {M\"unzenberg}}]{Kampfrath2013}%
  \BibitemOpen
  \bibfield  {author} {\bibinfo {author} {\bibfnamefont {T.}~\bibnamefont
  {Kampfrath}}, \bibinfo {author} {\bibfnamefont {M.}~\bibnamefont {Battiato}},
  \bibinfo {author} {\bibfnamefont {P.}~\bibnamefont {Maldonado}}, \bibinfo
  {author} {\bibfnamefont {G.}~\bibnamefont {Eilers}}, \bibinfo {author}
  {\bibfnamefont {J.}~\bibnamefont {N\"otzold}}, \bibinfo {author}
  {\bibfnamefont {S.}~\bibnamefont {M\"ahrlein}}, \bibinfo {author}
  {\bibfnamefont {V.}~\bibnamefont {Zbarsky}}, \bibinfo {author} {\bibfnamefont
  {F.}~\bibnamefont {Freimuth}}, \bibinfo {author} {\bibfnamefont
  {Y.}~\bibnamefont {Mokrousov}}, \bibinfo {author} {\bibfnamefont
  {S.}~\bibnamefont {Bl\"ugel}}, \bibinfo {author} {\bibfnamefont
  {M.}~\bibnamefont {Wolf}}, \bibinfo {author} {\bibfnamefont {I.}~\bibnamefont
  {Radu}}, \bibinfo {author} {\bibfnamefont {P.~M.}\ \bibnamefont {Oppeneer}},\
  and\ \bibinfo {author} {\bibfnamefont {M.}~\bibnamefont {M\"unzenberg}},\
  }\bibfield  {title} {\bibinfo {title} {Terahertz spin current pulses
  controlled by magnetic heterostructures},\ }\href
  {https://doi.org/10.1038/nnano.2013.43} {\bibfield  {journal} {\bibinfo
  {journal} {Nat. Nanotechnol.}\ }\textbf {\bibinfo {volume} {8}},\ \bibinfo
  {pages} {256} (\bibinfo {year} {2013})}\BibitemShut {NoStop}%
\bibitem [{\citenamefont {Seifert}\ \emph {et~al.}(2016)\citenamefont
  {Seifert}, \citenamefont {Jaiswal}, \citenamefont {Martens}, \citenamefont
  {Hannegan}, \citenamefont {Braun}, \citenamefont {Maldonado}, \citenamefont
  {Freimuth}, \citenamefont {Kronenberg}, \citenamefont {Henrizi},
  \citenamefont {Radu}, \citenamefont {Beaurepaire}, \citenamefont {Mokrousov},
  \citenamefont {Oppeneer}, \citenamefont {Jourdan}, \citenamefont {Jakob},
  \citenamefont {Turchinovich}, \citenamefont {Hayden}, \citenamefont {Wolf},
  \citenamefont {M\"unzenberg}, \citenamefont {Kl\"aui},\ and\ \citenamefont
  {Kampfrath}}]{Seifert2016}%
  \BibitemOpen
  \bibfield  {author} {\bibinfo {author} {\bibfnamefont {T.}~\bibnamefont
  {Seifert}}, \bibinfo {author} {\bibfnamefont {S.}~\bibnamefont {Jaiswal}},
  \bibinfo {author} {\bibfnamefont {U.}~\bibnamefont {Martens}}, \bibinfo
  {author} {\bibfnamefont {J.}~\bibnamefont {Hannegan}}, \bibinfo {author}
  {\bibfnamefont {L.}~\bibnamefont {Braun}}, \bibinfo {author} {\bibfnamefont
  {P.}~\bibnamefont {Maldonado}}, \bibinfo {author} {\bibfnamefont
  {F.}~\bibnamefont {Freimuth}}, \bibinfo {author} {\bibfnamefont
  {A.}~\bibnamefont {Kronenberg}}, \bibinfo {author} {\bibfnamefont
  {J.}~\bibnamefont {Henrizi}}, \bibinfo {author} {\bibfnamefont
  {I.}~\bibnamefont {Radu}}, \bibinfo {author} {\bibfnamefont {E.}~\bibnamefont
  {Beaurepaire}}, \bibinfo {author} {\bibfnamefont {Y.}~\bibnamefont
  {Mokrousov}}, \bibinfo {author} {\bibfnamefont {P.~M.}\ \bibnamefont
  {Oppeneer}}, \bibinfo {author} {\bibfnamefont {M.}~\bibnamefont {Jourdan}},
  \bibinfo {author} {\bibfnamefont {G.}~\bibnamefont {Jakob}}, \bibinfo
  {author} {\bibfnamefont {D.}~\bibnamefont {Turchinovich}}, \bibinfo {author}
  {\bibfnamefont {L.~M.}\ \bibnamefont {Hayden}}, \bibinfo {author}
  {\bibfnamefont {M.}~\bibnamefont {Wolf}}, \bibinfo {author} {\bibfnamefont
  {M.}~\bibnamefont {M\"unzenberg}}, \bibinfo {author} {\bibfnamefont
  {M.}~\bibnamefont {Kl\"aui}},\ and\ \bibinfo {author} {\bibfnamefont
  {T.}~\bibnamefont {Kampfrath}},\ }\bibfield  {title} {\bibinfo {title}
  {Efficient metallic spintronic emitters of ultrabroadband terahertz
  radiation},\ }\href {https://doi.org/10.1038/nphoton.2016.91} {\bibfield
  {journal} {\bibinfo  {journal} {Nat. Photonics}\ }\textbf {\bibinfo {volume}
  {10}},\ \bibinfo {pages} {483} (\bibinfo {year} {2016})}\BibitemShut
  {NoStop}%
\bibitem [{\citenamefont {Stanciu}\ \emph {et~al.}(2007)\citenamefont
  {Stanciu}, \citenamefont {Hansteen}, \citenamefont {Kimel}, \citenamefont
  {Kirilyuk}, \citenamefont {Tsukamoto}, \citenamefont {Itoh},\ and\
  \citenamefont {Rasing}}]{Stanciu2007}%
  \BibitemOpen
  \bibfield  {author} {\bibinfo {author} {\bibfnamefont {C.~D.}\ \bibnamefont
  {Stanciu}}, \bibinfo {author} {\bibfnamefont {F.}~\bibnamefont {Hansteen}},
  \bibinfo {author} {\bibfnamefont {A.~V.}\ \bibnamefont {Kimel}}, \bibinfo
  {author} {\bibfnamefont {A.}~\bibnamefont {Kirilyuk}}, \bibinfo {author}
  {\bibfnamefont {A.}~\bibnamefont {Tsukamoto}}, \bibinfo {author}
  {\bibfnamefont {A.}~\bibnamefont {Itoh}},\ and\ \bibinfo {author}
  {\bibfnamefont {T.}~\bibnamefont {Rasing}},\ }\bibfield  {title} {\bibinfo
  {title} {All-optical magnetic recording with circularly polarized light},\
  }\href {https://doi.org/10.1103/PhysRevLett.99.047601} {\bibfield  {journal}
  {\bibinfo  {journal} {Phys. Rev. Lett.}\ }\textbf {\bibinfo {volume} {99}},\
  \bibinfo {pages} {047601} (\bibinfo {year} {2007})}\BibitemShut {NoStop}%
\bibitem [{\citenamefont {Walowski}\ and\ \citenamefont
  {Münzenberg}(2016)}]{Walowski2016}%
  \BibitemOpen
  \bibfield  {author} {\bibinfo {author} {\bibfnamefont {J.}~\bibnamefont
  {Walowski}}\ and\ \bibinfo {author} {\bibfnamefont {M.}~\bibnamefont
  {Münzenberg}},\ }\bibfield  {title} {\bibinfo {title} {{Perspective:
  Ultrafast magnetism and THz spintronics}},\ }\href
  {https://doi.org/10.1063/1.4958846} {\bibfield  {journal} {\bibinfo
  {journal} {J. Appl. Phys.}\ }\textbf {\bibinfo {volume} {120}},\ \bibinfo
  {pages} {140901} (\bibinfo {year} {2016})}\BibitemShut {NoStop}%
\bibitem [{\citenamefont {Malinowski}\ \emph {et~al.}(2018)\citenamefont
  {Malinowski}, \citenamefont {Bergeard}, \citenamefont {Hehn},\ and\
  \citenamefont {Mangin}}]{Malinowski2018}%
  \BibitemOpen
  \bibfield  {author} {\bibinfo {author} {\bibfnamefont {G.}~\bibnamefont
  {Malinowski}}, \bibinfo {author} {\bibfnamefont {N.}~\bibnamefont
  {Bergeard}}, \bibinfo {author} {\bibfnamefont {M.}~\bibnamefont {Hehn}},\
  and\ \bibinfo {author} {\bibfnamefont {S.}~\bibnamefont {Mangin}},\
  }\bibfield  {title} {\bibinfo {title} {Hot-electron transport and ultrafast
  magnetization dynamics in magnetic multilayers and nanostructures following
  femtosecond laser pulse excitation},\ }\href
  {https://doi.org/10.1140/epjb/e2018-80555-5} {\bibfield  {journal} {\bibinfo
  {journal} {Eur. Phys. J. B}\ }\textbf {\bibinfo {volume} {91}},\ \bibinfo
  {pages} {98} (\bibinfo {year} {2018})}\BibitemShut {NoStop}%
\bibitem [{\citenamefont {Eschenlohr}\ \emph {et~al.}(2013)\citenamefont
  {Eschenlohr}, \citenamefont {Battiato}, \citenamefont {Maldonado},
  \citenamefont {Pontius}, \citenamefont {Kachel}, \citenamefont {Holldack},
  \citenamefont {Mitzner}, \citenamefont {F{\"o}hlisch}, \citenamefont
  {Oppeneer},\ and\ \citenamefont {Stamm}}]{Eschenlohr2013}%
  \BibitemOpen
  \bibfield  {author} {\bibinfo {author} {\bibfnamefont {A.}~\bibnamefont
  {Eschenlohr}}, \bibinfo {author} {\bibfnamefont {M.}~\bibnamefont
  {Battiato}}, \bibinfo {author} {\bibfnamefont {P.}~\bibnamefont {Maldonado}},
  \bibinfo {author} {\bibfnamefont {N.}~\bibnamefont {Pontius}}, \bibinfo
  {author} {\bibfnamefont {T.}~\bibnamefont {Kachel}}, \bibinfo {author}
  {\bibfnamefont {K.}~\bibnamefont {Holldack}}, \bibinfo {author}
  {\bibfnamefont {R.}~\bibnamefont {Mitzner}}, \bibinfo {author} {\bibfnamefont
  {A.}~\bibnamefont {F{\"o}hlisch}}, \bibinfo {author} {\bibfnamefont {P.~M.}\
  \bibnamefont {Oppeneer}},\ and\ \bibinfo {author} {\bibfnamefont
  {C.}~\bibnamefont {Stamm}},\ }\bibfield  {title} {\bibinfo {title} {Ultrafast
  spin transport as key to femtosecond demagnetization},\ }\href
  {https://doi.org/10.1038/nmat3546} {\bibfield  {journal} {\bibinfo  {journal}
  {Nat. Mater.}\ }\textbf {\bibinfo {volume} {12}},\ \bibinfo {pages} {332}
  (\bibinfo {year} {2013})}\BibitemShut {NoStop}%
\bibitem [{\citenamefont {Bergeard}\ \emph {et~al.}(2016)\citenamefont
  {Bergeard}, \citenamefont {Hehn}, \citenamefont {Mangin}, \citenamefont
  {Lengaigne}, \citenamefont {Montaigne}, \citenamefont {Lalieu}, \citenamefont
  {Koopmans},\ and\ \citenamefont {Malinowski}}]{Bergeard2016}%
  \BibitemOpen
  \bibfield  {author} {\bibinfo {author} {\bibfnamefont {N.}~\bibnamefont
  {Bergeard}}, \bibinfo {author} {\bibfnamefont {M.}~\bibnamefont {Hehn}},
  \bibinfo {author} {\bibfnamefont {S.}~\bibnamefont {Mangin}}, \bibinfo
  {author} {\bibfnamefont {G.}~\bibnamefont {Lengaigne}}, \bibinfo {author}
  {\bibfnamefont {F.}~\bibnamefont {Montaigne}}, \bibinfo {author}
  {\bibfnamefont {M.~L.~M.}\ \bibnamefont {Lalieu}}, \bibinfo {author}
  {\bibfnamefont {B.}~\bibnamefont {Koopmans}},\ and\ \bibinfo {author}
  {\bibfnamefont {G.}~\bibnamefont {Malinowski}},\ }\bibfield  {title}
  {\bibinfo {title} {Hot-electron-induced ultrafast demagnetization in
  $\mathrm{Co}/\mathrm{Pt}$ multilayers},\ }\href
  {https://doi.org/10.1103/PhysRevLett.117.147203} {\bibfield  {journal}
  {\bibinfo  {journal} {Phys. Rev. Lett.}\ }\textbf {\bibinfo {volume} {117}},\
  \bibinfo {pages} {147203} (\bibinfo {year} {2016})}\BibitemShut {NoStop}%
\bibitem [{\citenamefont {Battiato}\ \emph {et~al.}(2010)\citenamefont
  {Battiato}, \citenamefont {Carva},\ and\ \citenamefont
  {Oppeneer}}]{Battiato2010}%
  \BibitemOpen
  \bibfield  {author} {\bibinfo {author} {\bibfnamefont {M.}~\bibnamefont
  {Battiato}}, \bibinfo {author} {\bibfnamefont {K.}~\bibnamefont {Carva}},\
  and\ \bibinfo {author} {\bibfnamefont {P.~M.}\ \bibnamefont {Oppeneer}},\
  }\bibfield  {title} {\bibinfo {title} {Superdiffusive spin transport as a
  mechanism of ultrafast demagnetization},\ }\href
  {https://doi.org/10.1103/PhysRevLett.105.027203} {\bibfield  {journal}
  {\bibinfo  {journal} {Phys. Rev. Lett.}\ }\textbf {\bibinfo {volume} {105}},\
  \bibinfo {pages} {027203} (\bibinfo {year} {2010})}\BibitemShut {NoStop}%
\bibitem [{\citenamefont {Battiato}\ \emph {et~al.}(2012)\citenamefont
  {Battiato}, \citenamefont {Carva},\ and\ \citenamefont
  {Oppeneer}}]{Battiato2012}%
  \BibitemOpen
  \bibfield  {author} {\bibinfo {author} {\bibfnamefont {M.}~\bibnamefont
  {Battiato}}, \bibinfo {author} {\bibfnamefont {K.}~\bibnamefont {Carva}},\
  and\ \bibinfo {author} {\bibfnamefont {P.~M.}\ \bibnamefont {Oppeneer}},\
  }\bibfield  {title} {\bibinfo {title} {Theory of laser-induced ultrafast
  superdiffusive spin transport in layered heterostructures},\ }\href
  {https://doi.org/10.1103/PhysRevB.86.024404} {\bibfield  {journal} {\bibinfo
  {journal} {Phys. Rev. B}\ }\textbf {\bibinfo {volume} {86}},\ \bibinfo
  {pages} {024404} (\bibinfo {year} {2012})}\BibitemShut {NoStop}%
\bibitem [{\citenamefont {Bal{\'a}{\v z}}\ \emph {et~al.}(2018)\citenamefont
  {Bal{\'a}{\v z}}, \citenamefont {{\v Z}onda}, \citenamefont {Carva},
  \citenamefont {Maldonado},\ and\ \citenamefont {Oppeneer}}]{Balaz2018}%
  \BibitemOpen
  \bibfield  {author} {\bibinfo {author} {\bibfnamefont {P.}~\bibnamefont
  {Bal{\'a}{\v z}}}, \bibinfo {author} {\bibfnamefont {M.}~\bibnamefont {{\v
  Z}onda}}, \bibinfo {author} {\bibfnamefont {K.}~\bibnamefont {Carva}},
  \bibinfo {author} {\bibfnamefont {P.}~\bibnamefont {Maldonado}},\ and\
  \bibinfo {author} {\bibfnamefont {P.~M.}\ \bibnamefont {Oppeneer}},\
  }\bibfield  {title} {\bibinfo {title} {Transport theory for femtosecond
  laser-induced spin-transfer torques},\ }\href
  {https://doi.org/10.1088/1361-648X/aaad95} {\bibfield  {journal} {\bibinfo
  {journal} {J. Phys.: Condens. Matter}\ }\textbf {\bibinfo {volume} {30}},\
  \bibinfo {pages} {115801} (\bibinfo {year} {2018})}\BibitemShut {NoStop}%
\bibitem [{\citenamefont {Razdolski}\ \emph {et~al.}(2017)\citenamefont
  {Razdolski}, \citenamefont {Alekhin}, \citenamefont {Ilin}, \citenamefont
  {Meyburg}, \citenamefont {Roddatis}, \citenamefont {Diesing}, \citenamefont
  {Bovensiepen},\ and\ \citenamefont {Melnikov}}]{Razdolski2017}%
  \BibitemOpen
  \bibfield  {author} {\bibinfo {author} {\bibfnamefont {I.}~\bibnamefont
  {Razdolski}}, \bibinfo {author} {\bibfnamefont {A.}~\bibnamefont {Alekhin}},
  \bibinfo {author} {\bibfnamefont {N.}~\bibnamefont {Ilin}}, \bibinfo {author}
  {\bibfnamefont {J.~P.}\ \bibnamefont {Meyburg}}, \bibinfo {author}
  {\bibfnamefont {V.}~\bibnamefont {Roddatis}}, \bibinfo {author}
  {\bibfnamefont {D.}~\bibnamefont {Diesing}}, \bibinfo {author} {\bibfnamefont
  {U.}~\bibnamefont {Bovensiepen}},\ and\ \bibinfo {author} {\bibfnamefont
  {A.}~\bibnamefont {Melnikov}},\ }\bibfield  {title} {\bibinfo {title}
  {Nanoscale interface confinement of ultrafast spin transfer torque driving
  non-uniform spin dynamics},\ }\href {https://doi.org/10.1038/ncomms15007}
  {\bibfield  {journal} {\bibinfo  {journal} {Nat. Commun.}\ }\textbf {\bibinfo
  {volume} {8}},\ \bibinfo {pages} {15007} (\bibinfo {year}
  {2017})}\BibitemShut {NoStop}%
\bibitem [{\citenamefont {Ulrichs}\ and\ \citenamefont
  {Razdolski}(2018)}]{Ulrichs2018}%
  \BibitemOpen
  \bibfield  {author} {\bibinfo {author} {\bibfnamefont {H.}~\bibnamefont
  {Ulrichs}}\ and\ \bibinfo {author} {\bibfnamefont {I.}~\bibnamefont
  {Razdolski}},\ }\bibfield  {title} {\bibinfo {title} {Micromagnetic view on
  ultrafast magnon generation by femtosecond spin current pulses},\ }\href
  {https://doi.org/10.1103/PhysRevB.98.054429} {\bibfield  {journal} {\bibinfo
  {journal} {Phys. Rev. B}\ }\textbf {\bibinfo {volume} {98}},\ \bibinfo
  {pages} {054429} (\bibinfo {year} {2018})}\BibitemShut {NoStop}%
\bibitem [{\citenamefont {Ritzmann}\ \emph {et~al.}(2020)\citenamefont
  {Ritzmann}, \citenamefont {Bal\'a\ifmmode~\check{z}\else \v{z}\fi{}},
  \citenamefont {Maldonado}, \citenamefont {Carva},\ and\ \citenamefont
  {Oppeneer}}]{Ritzmann2020}%
  \BibitemOpen
  \bibfield  {author} {\bibinfo {author} {\bibfnamefont {U.}~\bibnamefont
  {Ritzmann}}, \bibinfo {author} {\bibfnamefont {P.}~\bibnamefont
  {Bal\'a\ifmmode~\check{z}\else \v{z}\fi{}}}, \bibinfo {author} {\bibfnamefont
  {P.}~\bibnamefont {Maldonado}}, \bibinfo {author} {\bibfnamefont
  {K.}~\bibnamefont {Carva}},\ and\ \bibinfo {author} {\bibfnamefont {P.~M.}\
  \bibnamefont {Oppeneer}},\ }\bibfield  {title} {\bibinfo {title}
  {High-frequency magnon excitation due to femtosecond spin-transfer torques},\
  }\href {https://doi.org/10.1103/PhysRevB.101.174427} {\bibfield  {journal}
  {\bibinfo  {journal} {Phys. Rev. B}\ }\textbf {\bibinfo {volume} {101}},\
  \bibinfo {pages} {174427} (\bibinfo {year} {2020})}\BibitemShut {NoStop}%
\bibitem [{\citenamefont {Jungwirth}\ \emph {et~al.}(2016)\citenamefont
  {Jungwirth}, \citenamefont {Marti}, \citenamefont {Wadley},\ and\
  \citenamefont {Wunderlich}}]{Jungwirth2016}%
  \BibitemOpen
  \bibfield  {author} {\bibinfo {author} {\bibfnamefont {T.}~\bibnamefont
  {Jungwirth}}, \bibinfo {author} {\bibfnamefont {X.}~\bibnamefont {Marti}},
  \bibinfo {author} {\bibfnamefont {P.}~\bibnamefont {Wadley}},\ and\ \bibinfo
  {author} {\bibfnamefont {J.}~\bibnamefont {Wunderlich}},\ }\bibfield  {title}
  {\bibinfo {title} {Antiferromagnetic spintronics},\ }\href
  {https://doi.org/10.1038/nnano.2016.18} {\bibfield  {journal} {\bibinfo
  {journal} {Nat. Nanotechnol.}\ }\textbf {\bibinfo {volume} {11}},\ \bibinfo
  {pages} {231} (\bibinfo {year} {2016})}\BibitemShut {NoStop}%
\bibitem [{\citenamefont {Baltz}\ \emph {et~al.}(2018)\citenamefont {Baltz},
  \citenamefont {Manchon}, \citenamefont {Tsoi}, \citenamefont {Moriyama},
  \citenamefont {Ono},\ and\ \citenamefont {Tserkovnyak}}]{Baltz2018}%
  \BibitemOpen
  \bibfield  {author} {\bibinfo {author} {\bibfnamefont {V.}~\bibnamefont
  {Baltz}}, \bibinfo {author} {\bibfnamefont {A.}~\bibnamefont {Manchon}},
  \bibinfo {author} {\bibfnamefont {M.}~\bibnamefont {Tsoi}}, \bibinfo {author}
  {\bibfnamefont {T.}~\bibnamefont {Moriyama}}, \bibinfo {author}
  {\bibfnamefont {T.}~\bibnamefont {Ono}},\ and\ \bibinfo {author}
  {\bibfnamefont {Y.}~\bibnamefont {Tserkovnyak}},\ }\bibfield  {title}
  {\bibinfo {title} {Antiferromagnetic spintronics},\ }\href
  {https://doi.org/10.1103/RevModPhys.90.015005} {\bibfield  {journal}
  {\bibinfo  {journal} {Rev. Mod. Phys.}\ }\textbf {\bibinfo {volume} {90}},\
  \bibinfo {pages} {015005} (\bibinfo {year} {2018})}\BibitemShut {NoStop}%
\bibitem [{\citenamefont {Cheng}\ and\ \citenamefont {Niu}(2014)}]{Cheng2014}%
  \BibitemOpen
  \bibfield  {author} {\bibinfo {author} {\bibfnamefont {R.}~\bibnamefont
  {Cheng}}\ and\ \bibinfo {author} {\bibfnamefont {Q.}~\bibnamefont {Niu}},\
  }\bibfield  {title} {\bibinfo {title} {Dynamics of antiferromagnets driven by
  spin current},\ }\href {https://doi.org/10.1103/PhysRevB.89.081105}
  {\bibfield  {journal} {\bibinfo  {journal} {Phys. Rev. B}\ }\textbf {\bibinfo
  {volume} {89}},\ \bibinfo {pages} {081105} (\bibinfo {year}
  {2014})}\BibitemShut {NoStop}%
\bibitem [{\citenamefont {Kimel}\ \emph {et~al.}(2009)\citenamefont {Kimel},
  \citenamefont {Ivanov}, \citenamefont {Pisarev}, \citenamefont {Usachev},
  \citenamefont {Kirilyuk},\ and\ \citenamefont {Rasing}}]{Kimel2009}%
  \BibitemOpen
  \bibfield  {author} {\bibinfo {author} {\bibfnamefont {A.~V.}\ \bibnamefont
  {Kimel}}, \bibinfo {author} {\bibfnamefont {B.~A.}\ \bibnamefont {Ivanov}},
  \bibinfo {author} {\bibfnamefont {R.~V.}\ \bibnamefont {Pisarev}}, \bibinfo
  {author} {\bibfnamefont {P.~A.}\ \bibnamefont {Usachev}}, \bibinfo {author}
  {\bibfnamefont {A.}~\bibnamefont {Kirilyuk}},\ and\ \bibinfo {author}
  {\bibfnamefont {T.}~\bibnamefont {Rasing}},\ }\bibfield  {title} {\bibinfo
  {title} {Inertia-driven spin switching in antiferromagnets},\ }\href
  {https://doi.org/10.1038/nphys1369} {\bibfield  {journal} {\bibinfo
  {journal} {Nat. Phys.}\ }\textbf {\bibinfo {volume} {5}},\ \bibinfo {pages}
  {727} (\bibinfo {year} {2009})}\BibitemShut {NoStop}%
\bibitem [{\citenamefont {Kampfrath}\ \emph {et~al.}(2011)\citenamefont
  {Kampfrath}, \citenamefont {Sell}, \citenamefont {Klatt}, \citenamefont
  {Pashkin}, \citenamefont {M{\"a}hrlein}, \citenamefont {Dekorsy},
  \citenamefont {Wolf}, \citenamefont {Fiebig}, \citenamefont {Leitenstorfer},\
  and\ \citenamefont {Huber}}]{Kampfrath2011}%
  \BibitemOpen
  \bibfield  {author} {\bibinfo {author} {\bibfnamefont {T.}~\bibnamefont
  {Kampfrath}}, \bibinfo {author} {\bibfnamefont {A.}~\bibnamefont {Sell}},
  \bibinfo {author} {\bibfnamefont {G.}~\bibnamefont {Klatt}}, \bibinfo
  {author} {\bibfnamefont {A.}~\bibnamefont {Pashkin}}, \bibinfo {author}
  {\bibfnamefont {S.}~\bibnamefont {M{\"a}hrlein}}, \bibinfo {author}
  {\bibfnamefont {T.}~\bibnamefont {Dekorsy}}, \bibinfo {author} {\bibfnamefont
  {M.}~\bibnamefont {Wolf}}, \bibinfo {author} {\bibfnamefont {M.}~\bibnamefont
  {Fiebig}}, \bibinfo {author} {\bibfnamefont {A.}~\bibnamefont
  {Leitenstorfer}},\ and\ \bibinfo {author} {\bibfnamefont {R.}~\bibnamefont
  {Huber}},\ }\bibfield  {title} {\bibinfo {title} {Coherent terahertz control
  of antiferromagnetic spin waves},\ }\href
  {https://doi.org/10.1038/nphoton.2010.259} {\bibfield  {journal} {\bibinfo
  {journal} {Nat. Photonics}\ }\textbf {\bibinfo {volume} {5}},\ \bibinfo
  {pages} {31} (\bibinfo {year} {2011})}\BibitemShut {NoStop}%
\bibitem [{\citenamefont {Hortensius}\ \emph {et~al.}(2021)\citenamefont
  {Hortensius}, \citenamefont {Afanasiev}, \citenamefont {Matthiesen},
  \citenamefont {Leenders}, \citenamefont {Citro}, \citenamefont {Kimel},
  \citenamefont {Mikhaylovskiy}, \citenamefont {Ivanov},\ and\ \citenamefont
  {Caviglia}}]{Hortensius2021}%
  \BibitemOpen
  \bibfield  {author} {\bibinfo {author} {\bibfnamefont {J.~R.}\ \bibnamefont
  {Hortensius}}, \bibinfo {author} {\bibfnamefont {D.}~\bibnamefont
  {Afanasiev}}, \bibinfo {author} {\bibfnamefont {M.}~\bibnamefont
  {Matthiesen}}, \bibinfo {author} {\bibfnamefont {R.}~\bibnamefont
  {Leenders}}, \bibinfo {author} {\bibfnamefont {R.}~\bibnamefont {Citro}},
  \bibinfo {author} {\bibfnamefont {A.~V.}\ \bibnamefont {Kimel}}, \bibinfo
  {author} {\bibfnamefont {R.~V.}\ \bibnamefont {Mikhaylovskiy}}, \bibinfo
  {author} {\bibfnamefont {B.~A.}\ \bibnamefont {Ivanov}},\ and\ \bibinfo
  {author} {\bibfnamefont {A.~D.}\ \bibnamefont {Caviglia}},\ }\bibfield
  {title} {\bibinfo {title} {Coherent spin-wave transport in an
  antiferromagnet},\ }\href {https://doi.org/10.1038/s41567-021-01290-4}
  {\bibfield  {journal} {\bibinfo  {journal} {Nat. Phys.}\ }\textbf {\bibinfo
  {volume} {17}},\ \bibinfo {pages} {1001} (\bibinfo {year}
  {2021})}\BibitemShut {NoStop}%
\bibitem [{\citenamefont {Ross}\ \emph {et~al.}(2024)\citenamefont {Ross},
  \citenamefont {Gavriloaea}, \citenamefont {Freimuth}, \citenamefont
  {Adamantopoulos}, \citenamefont {Mokrousov}, \citenamefont {Evans},
  \citenamefont {Chantrell}, \citenamefont {Otxoa},\ and\ \citenamefont
  {Chubykalo-Fesenko}}]{Ross2024}%
  \BibitemOpen
  \bibfield  {author} {\bibinfo {author} {\bibfnamefont {J.~L.}\ \bibnamefont
  {Ross}}, \bibinfo {author} {\bibfnamefont {P.-I.}\ \bibnamefont
  {Gavriloaea}}, \bibinfo {author} {\bibfnamefont {F.}~\bibnamefont
  {Freimuth}}, \bibinfo {author} {\bibfnamefont {T.}~\bibnamefont
  {Adamantopoulos}}, \bibinfo {author} {\bibfnamefont {Y.}~\bibnamefont
  {Mokrousov}}, \bibinfo {author} {\bibfnamefont {R.~F.~L.}\ \bibnamefont
  {Evans}}, \bibinfo {author} {\bibfnamefont {R.}~\bibnamefont {Chantrell}},
  \bibinfo {author} {\bibfnamefont {R.~M.}\ \bibnamefont {Otxoa}},\ and\
  \bibinfo {author} {\bibfnamefont {O.}~\bibnamefont {Chubykalo-Fesenko}},\
  }\bibfield  {title} {\bibinfo {title} {Ultrafast antiferromagnetic switching
  of {Mn}$_2${Au} with laser-induced optical torques},\ }\href
  {https://doi.org/10.1038/s41524-024-01416-1} {\bibfield  {journal} {\bibinfo
  {journal} {npj Comput. Mater.}\ }\textbf {\bibinfo {volume} {10}},\ \bibinfo
  {pages} {234} (\bibinfo {year} {2024})}\BibitemShut {NoStop}%
\bibitem [{\citenamefont {Chirac}\ \emph {et~al.}(2020)\citenamefont {Chirac},
  \citenamefont {Chauleau}, \citenamefont {Thibaudeau}, \citenamefont
  {Gomonay},\ and\ \citenamefont {Viret}}]{Chirac2020}%
  \BibitemOpen
  \bibfield  {author} {\bibinfo {author} {\bibfnamefont {T.}~\bibnamefont
  {Chirac}}, \bibinfo {author} {\bibfnamefont {J.-Y.}\ \bibnamefont
  {Chauleau}}, \bibinfo {author} {\bibfnamefont {P.}~\bibnamefont
  {Thibaudeau}}, \bibinfo {author} {\bibfnamefont {O.}~\bibnamefont
  {Gomonay}},\ and\ \bibinfo {author} {\bibfnamefont {M.}~\bibnamefont
  {Viret}},\ }\bibfield  {title} {\bibinfo {title} {Ultrafast antiferromagnetic
  switching in nio induced by spin transfer torques},\ }\href
  {https://doi.org/10.1103/PhysRevB.102.134415} {\bibfield  {journal} {\bibinfo
   {journal} {Phys. Rev. B}\ }\textbf {\bibinfo {volume} {102}},\ \bibinfo
  {pages} {134415} (\bibinfo {year} {2020})}\BibitemShut {NoStop}%
\bibitem [{\citenamefont {Wei\ss{}enhofer}\ \emph {et~al.}(2023)\citenamefont
  {Wei\ss{}enhofer}, \citenamefont {Foggetti}, \citenamefont {Nowak},\ and\
  \citenamefont {Oppeneer}}]{Weibenhofer2023}%
  \BibitemOpen
  \bibfield  {author} {\bibinfo {author} {\bibfnamefont {M.}~\bibnamefont
  {Wei\ss{}enhofer}}, \bibinfo {author} {\bibfnamefont {F.}~\bibnamefont
  {Foggetti}}, \bibinfo {author} {\bibfnamefont {U.}~\bibnamefont {Nowak}},\
  and\ \bibinfo {author} {\bibfnamefont {P.~M.}\ \bibnamefont {Oppeneer}},\
  }\bibfield  {title} {\bibinfo {title} {N\'eel vector switching and terahertz
  spin-wave excitation in {Mn}$_2${Au} due to femtosecond spin-transfer
  torques},\ }\href {https://doi.org/10.1103/PhysRevB.107.174424} {\bibfield
  {journal} {\bibinfo  {journal} {Phys. Rev. B}\ }\textbf {\bibinfo {volume}
  {107}},\ \bibinfo {pages} {174424} (\bibinfo {year} {2023})}\BibitemShut
  {NoStop}%
\bibitem [{\citenamefont {Chen}\ \emph {et~al.}(2014)\citenamefont {Chen},
  \citenamefont {Niu},\ and\ \citenamefont {MacDonald}}]{Chen2014}%
  \BibitemOpen
  \bibfield  {author} {\bibinfo {author} {\bibfnamefont {H.}~\bibnamefont
  {Chen}}, \bibinfo {author} {\bibfnamefont {Q.}~\bibnamefont {Niu}},\ and\
  \bibinfo {author} {\bibfnamefont {A.~H.}\ \bibnamefont {MacDonald}},\
  }\bibfield  {title} {\bibinfo {title} {Anomalous {Hall} effect arising from
  noncollinear antiferromagnetism},\ }\href
  {https://doi.org/10.1103/PhysRevLett.112.017205} {\bibfield  {journal}
  {\bibinfo  {journal} {Phys. Rev. Lett.}\ }\textbf {\bibinfo {volume} {112}},\
  \bibinfo {pages} {017205} (\bibinfo {year} {2014})}\BibitemShut {NoStop}%
\bibitem [{\citenamefont {K{\"u}bler}\ and\ \citenamefont
  {Felser}(2014)}]{Kubler_2014}%
  \BibitemOpen
  \bibfield  {author} {\bibinfo {author} {\bibfnamefont {J.}~\bibnamefont
  {K{\"u}bler}}\ and\ \bibinfo {author} {\bibfnamefont {C.}~\bibnamefont
  {Felser}},\ }\bibfield  {title} {\bibinfo {title} {Non-collinear
  antiferromagnets and the anomalous {Hall} effect},\ }\href
  {https://doi.org/10.1209/0295-5075/108/67001} {\bibfield  {journal} {\bibinfo
   {journal} {Europhys. Lett.}\ }\textbf {\bibinfo {volume} {108}},\ \bibinfo
  {pages} {67001} (\bibinfo {year} {2014})}\BibitemShut {NoStop}%
\bibitem [{\citenamefont {Nakatsuji}\ \emph {et~al.}(2015)\citenamefont
  {Nakatsuji}, \citenamefont {Kiyohara},\ and\ \citenamefont
  {Higo}}]{Nakatsuji2015}%
  \BibitemOpen
  \bibfield  {author} {\bibinfo {author} {\bibfnamefont {S.}~\bibnamefont
  {Nakatsuji}}, \bibinfo {author} {\bibfnamefont {N.}~\bibnamefont
  {Kiyohara}},\ and\ \bibinfo {author} {\bibfnamefont {T.}~\bibnamefont
  {Higo}},\ }\bibfield  {title} {\bibinfo {title} {Large anomalous {Hall}
  effect in a non-collinear antiferromagnet at room temperature},\ }\href
  {https://doi.org/10.1038/nature15723} {\bibfield  {journal} {\bibinfo
  {journal} {Nature (London)}\ }\textbf {\bibinfo {volume} {527}},\ \bibinfo
  {pages} {212} (\bibinfo {year} {2015})}\BibitemShut {NoStop}%
\bibitem [{\citenamefont {Nayak}\ \emph {et~al.}(2016)\citenamefont {Nayak},
  \citenamefont {Fischer}, \citenamefont {Sun}, \citenamefont {Yan},
  \citenamefont {Karel}, \citenamefont {Komarek}, \citenamefont {Shekhar},
  \citenamefont {Kumar}, \citenamefont {Schnelle}, \citenamefont {K{\"u}bler},
  \citenamefont {Felser},\ and\ \citenamefont {Parkin}}]{Ajaya2016}%
  \BibitemOpen
  \bibfield  {author} {\bibinfo {author} {\bibfnamefont {A.~K.}\ \bibnamefont
  {Nayak}}, \bibinfo {author} {\bibfnamefont {J.~E.}\ \bibnamefont {Fischer}},
  \bibinfo {author} {\bibfnamefont {Y.}~\bibnamefont {Sun}}, \bibinfo {author}
  {\bibfnamefont {B.}~\bibnamefont {Yan}}, \bibinfo {author} {\bibfnamefont
  {J.}~\bibnamefont {Karel}}, \bibinfo {author} {\bibfnamefont {A.~C.}\
  \bibnamefont {Komarek}}, \bibinfo {author} {\bibfnamefont {C.}~\bibnamefont
  {Shekhar}}, \bibinfo {author} {\bibfnamefont {N.}~\bibnamefont {Kumar}},
  \bibinfo {author} {\bibfnamefont {W.}~\bibnamefont {Schnelle}}, \bibinfo
  {author} {\bibfnamefont {J.}~\bibnamefont {K{\"u}bler}}, \bibinfo {author}
  {\bibfnamefont {C.}~\bibnamefont {Felser}},\ and\ \bibinfo {author}
  {\bibfnamefont {S.~S.~P.}\ \bibnamefont {Parkin}},\ }\bibfield  {title}
  {\bibinfo {title} {Large anomalous {Hall} effect driven by a nonvanishing
  {Berry} curvature in the noncolinear antiferromagnet {Mn}$_3${Ge}},\ }\href
  {https://doi.org/10.1126/sciadv.1501870} {\bibfield  {journal} {\bibinfo
  {journal} {Sci. Adv.}\ }\textbf {\bibinfo {volume} {2}},\ \bibinfo {pages}
  {e1501870} (\bibinfo {year} {2016})}\BibitemShut {NoStop}%
\bibitem [{\citenamefont {Ikhlas}\ \emph {et~al.}(2017)\citenamefont {Ikhlas},
  \citenamefont {Tomita}, \citenamefont {Koretsune}, \citenamefont {Suzuki},
  \citenamefont {Nishio-Hamane}, \citenamefont {Arita}, \citenamefont {Otani},\
  and\ \citenamefont {Nakatsuji}}]{Ikhlas2017}%
  \BibitemOpen
  \bibfield  {author} {\bibinfo {author} {\bibfnamefont {M.}~\bibnamefont
  {Ikhlas}}, \bibinfo {author} {\bibfnamefont {T.}~\bibnamefont {Tomita}},
  \bibinfo {author} {\bibfnamefont {T.}~\bibnamefont {Koretsune}}, \bibinfo
  {author} {\bibfnamefont {M.-T.}\ \bibnamefont {Suzuki}}, \bibinfo {author}
  {\bibfnamefont {D.}~\bibnamefont {Nishio-Hamane}}, \bibinfo {author}
  {\bibfnamefont {R.}~\bibnamefont {Arita}}, \bibinfo {author} {\bibfnamefont
  {Y.}~\bibnamefont {Otani}},\ and\ \bibinfo {author} {\bibfnamefont
  {S.}~\bibnamefont {Nakatsuji}},\ }\bibfield  {title} {\bibinfo {title} {Large
  anomalous {Nernst} effect at room temperature in a chiral antiferromagnet},\
  }\href {https://doi.org/10.1038/nphys4181} {\bibfield  {journal} {\bibinfo
  {journal} {Nat. Phys.}\ }\textbf {\bibinfo {volume} {13}},\ \bibinfo {pages}
  {1085} (\bibinfo {year} {2017})}\BibitemShut {NoStop}%
\bibitem [{\citenamefont {Reichlova}\ \emph {et~al.}(2019)\citenamefont
  {Reichlova}, \citenamefont {Janda}, \citenamefont {Godinho}, \citenamefont
  {Markou}, \citenamefont {Kriegner}, \citenamefont {Schlitz}, \citenamefont
  {Zelezny}, \citenamefont {Soban}, \citenamefont {Bejarano}, \citenamefont
  {Schultheiss}, \citenamefont {Nemec}, \citenamefont {Jungwirth},
  \citenamefont {Felser}, \citenamefont {Wunderlich},\ and\ \citenamefont
  {Goennenwein}}]{Reichlova2019}%
  \BibitemOpen
  \bibfield  {author} {\bibinfo {author} {\bibfnamefont {H.}~\bibnamefont
  {Reichlova}}, \bibinfo {author} {\bibfnamefont {T.}~\bibnamefont {Janda}},
  \bibinfo {author} {\bibfnamefont {J.}~\bibnamefont {Godinho}}, \bibinfo
  {author} {\bibfnamefont {A.}~\bibnamefont {Markou}}, \bibinfo {author}
  {\bibfnamefont {D.}~\bibnamefont {Kriegner}}, \bibinfo {author}
  {\bibfnamefont {R.}~\bibnamefont {Schlitz}}, \bibinfo {author} {\bibfnamefont
  {J.}~\bibnamefont {Zelezny}}, \bibinfo {author} {\bibfnamefont
  {Z.}~\bibnamefont {Soban}}, \bibinfo {author} {\bibfnamefont
  {M.}~\bibnamefont {Bejarano}}, \bibinfo {author} {\bibfnamefont
  {H.}~\bibnamefont {Schultheiss}}, \bibinfo {author} {\bibfnamefont
  {P.}~\bibnamefont {Nemec}}, \bibinfo {author} {\bibfnamefont
  {T.}~\bibnamefont {Jungwirth}}, \bibinfo {author} {\bibfnamefont
  {C.}~\bibnamefont {Felser}}, \bibinfo {author} {\bibfnamefont
  {J.}~\bibnamefont {Wunderlich}},\ and\ \bibinfo {author} {\bibfnamefont
  {S.~T.~B.}\ \bibnamefont {Goennenwein}},\ }\bibfield  {title} {\bibinfo
  {title} {Imaging and writing magnetic domains in the non-collinear
  antiferromagnet {Mn$_3$Sn}},\ }\href
  {https://doi.org/10.1038/s41467-019-13391-z} {\bibfield  {journal} {\bibinfo
  {journal} {Nat. Commun.}\ }\textbf {\bibinfo {volume} {10}},\ \bibinfo
  {pages} {5459} (\bibinfo {year} {2019})}\BibitemShut {NoStop}%
\bibitem [{\citenamefont {Higo}\ \emph {et~al.}(2018)\citenamefont {Higo},
  \citenamefont {Man}, \citenamefont {Gopman}, \citenamefont {Wu},
  \citenamefont {Koretsune}, \citenamefont {van~’t Erve}, \citenamefont
  {Kabanov}, \citenamefont {Rees}, \citenamefont {Li}, \citenamefont {Suzuki},
  \citenamefont {Patankar}, \citenamefont {Ikhlas}, \citenamefont {Chien},
  \citenamefont {Arita}, \citenamefont {Shull}, \citenamefont {Orenstein},\
  and\ \citenamefont {Nakatsuji}}]{Higo2018}%
  \BibitemOpen
  \bibfield  {author} {\bibinfo {author} {\bibfnamefont {T.}~\bibnamefont
  {Higo}}, \bibinfo {author} {\bibfnamefont {H.}~\bibnamefont {Man}}, \bibinfo
  {author} {\bibfnamefont {D.~B.}\ \bibnamefont {Gopman}}, \bibinfo {author}
  {\bibfnamefont {L.}~\bibnamefont {Wu}}, \bibinfo {author} {\bibfnamefont
  {T.}~\bibnamefont {Koretsune}}, \bibinfo {author} {\bibfnamefont {O.~M.~J.}\
  \bibnamefont {van~’t Erve}}, \bibinfo {author} {\bibfnamefont {Y.~P.}\
  \bibnamefont {Kabanov}}, \bibinfo {author} {\bibfnamefont {D.}~\bibnamefont
  {Rees}}, \bibinfo {author} {\bibfnamefont {Y.}~\bibnamefont {Li}}, \bibinfo
  {author} {\bibfnamefont {M.-T.}\ \bibnamefont {Suzuki}}, \bibinfo {author}
  {\bibfnamefont {S.}~\bibnamefont {Patankar}}, \bibinfo {author}
  {\bibfnamefont {M.}~\bibnamefont {Ikhlas}}, \bibinfo {author} {\bibfnamefont
  {C.~L.}\ \bibnamefont {Chien}}, \bibinfo {author} {\bibfnamefont
  {R.}~\bibnamefont {Arita}}, \bibinfo {author} {\bibfnamefont {R.~D.}\
  \bibnamefont {Shull}}, \bibinfo {author} {\bibfnamefont {J.}~\bibnamefont
  {Orenstein}},\ and\ \bibinfo {author} {\bibfnamefont {S.}~\bibnamefont
  {Nakatsuji}},\ }\bibfield  {title} {\bibinfo {title} {Large magneto-optical
  {Kerr} effect and imaging of magnetic octupole domains in an
  antiferromagnetic metal},\ }\href {https://doi.org/10.1038/s41566-017-0086-z}
  {\bibfield  {journal} {\bibinfo  {journal} {Nat. Photonics}\ }\textbf
  {\bibinfo {volume} {12}},\ \bibinfo {pages} {73} (\bibinfo {year}
  {2018})}\BibitemShut {NoStop}%
\bibitem [{\citenamefont {Shukla}\ and\ \citenamefont
  {Rakheja}(2022)}]{Shukla2022}%
  \BibitemOpen
  \bibfield  {author} {\bibinfo {author} {\bibfnamefont {A.}~\bibnamefont
  {Shukla}}\ and\ \bibinfo {author} {\bibfnamefont {S.}~\bibnamefont
  {Rakheja}},\ }\bibfield  {title} {\bibinfo {title} {Spin-torque-driven
  terahertz auto-oscillations in noncollinear coplanar antiferromagnets},\
  }\href {https://doi.org/10.1103/PhysRevApplied.17.034037} {\bibfield
  {journal} {\bibinfo  {journal} {Phys. Rev. Appl.}\ }\textbf {\bibinfo
  {volume} {17}},\ \bibinfo {pages} {034037} (\bibinfo {year}
  {2022})}\BibitemShut {NoStop}%
\bibitem [{\citenamefont {Tsai}\ \emph {et~al.}(2020)\citenamefont {Tsai},
  \citenamefont {Higo}, \citenamefont {Kondou}, \citenamefont {Nomoto},
  \citenamefont {Sakai}, \citenamefont {Kobayashi}, \citenamefont {Nakano},
  \citenamefont {Yakushiji}, \citenamefont {Arita}, \citenamefont {Miwa},
  \citenamefont {Otani},\ and\ \citenamefont {Nakatsuji}}]{Tsai2020}%
  \BibitemOpen
  \bibfield  {author} {\bibinfo {author} {\bibfnamefont {H.}~\bibnamefont
  {Tsai}}, \bibinfo {author} {\bibfnamefont {T.}~\bibnamefont {Higo}}, \bibinfo
  {author} {\bibfnamefont {K.}~\bibnamefont {Kondou}}, \bibinfo {author}
  {\bibfnamefont {T.}~\bibnamefont {Nomoto}}, \bibinfo {author} {\bibfnamefont
  {A.}~\bibnamefont {Sakai}}, \bibinfo {author} {\bibfnamefont
  {A.}~\bibnamefont {Kobayashi}}, \bibinfo {author} {\bibfnamefont
  {T.}~\bibnamefont {Nakano}}, \bibinfo {author} {\bibfnamefont
  {K.}~\bibnamefont {Yakushiji}}, \bibinfo {author} {\bibfnamefont
  {R.}~\bibnamefont {Arita}}, \bibinfo {author} {\bibfnamefont
  {S.}~\bibnamefont {Miwa}}, \bibinfo {author} {\bibfnamefont {Y.}~\bibnamefont
  {Otani}},\ and\ \bibinfo {author} {\bibfnamefont {S.}~\bibnamefont
  {Nakatsuji}},\ }\bibfield  {title} {\bibinfo {title} {Electrical manipulation
  of a topological antiferromagnetic state},\ }\href
  {https://doi.org/10.1038/s41586-020-2211-2} {\bibfield  {journal} {\bibinfo
  {journal} {Nature (London)}\ }\textbf {\bibinfo {volume} {580}},\ \bibinfo
  {pages} {608} (\bibinfo {year} {2020})}\BibitemShut {NoStop}%
\bibitem [{\citenamefont {Kiyohara}\ \emph {et~al.}(2016)\citenamefont
  {Kiyohara}, \citenamefont {Tomita},\ and\ \citenamefont
  {Nakatsuji}}]{Kiyohara2016}%
  \BibitemOpen
  \bibfield  {author} {\bibinfo {author} {\bibfnamefont {N.}~\bibnamefont
  {Kiyohara}}, \bibinfo {author} {\bibfnamefont {T.}~\bibnamefont {Tomita}},\
  and\ \bibinfo {author} {\bibfnamefont {S.}~\bibnamefont {Nakatsuji}},\
  }\bibfield  {title} {\bibinfo {title} {Giant anomalous {Hall} effect in the
  chiral antiferromagnet {Mn$_3$Ge}},\ }\href
  {https://doi.org/10.1103/PhysRevApplied.5.064009} {\bibfield  {journal}
  {\bibinfo  {journal} {Phys. Rev. Appl.}\ }\textbf {\bibinfo {volume} {5}},\
  \bibinfo {pages} {064009} (\bibinfo {year} {2016})}\BibitemShut {NoStop}%
\bibitem [{\citenamefont {Nomoto}\ and\ \citenamefont
  {Arita}(2020)}]{Nomoto2020}%
  \BibitemOpen
  \bibfield  {author} {\bibinfo {author} {\bibfnamefont {T.}~\bibnamefont
  {Nomoto}}\ and\ \bibinfo {author} {\bibfnamefont {R.}~\bibnamefont {Arita}},\
  }\bibfield  {title} {\bibinfo {title} {Cluster multipole dynamics in
  noncollinear antiferromagnets},\ }\href
  {https://doi.org/10.1103/PhysRevResearch.2.012045} {\bibfield  {journal}
  {\bibinfo  {journal} {Phys. Rev. Res.}\ }\textbf {\bibinfo {volume} {2}},\
  \bibinfo {pages} {012045} (\bibinfo {year} {2020})}\BibitemShut {NoStop}%
\bibitem [{\citenamefont {Dasgupta}(2022)}]{Dasgupta2022}%
  \BibitemOpen
  \bibfield  {author} {\bibinfo {author} {\bibfnamefont {S.}~\bibnamefont
  {Dasgupta}},\ }\bibfield  {title} {\bibinfo {title} {Tuning the transport
  properties of {Mn}$_3${Ge} through the effect of strain on its magnetism},\
  }\href {https://doi.org/10.1103/PhysRevB.106.064431} {\bibfield  {journal}
  {\bibinfo  {journal} {Phys. Rev. B}\ }\textbf {\bibinfo {volume} {106}},\
  \bibinfo {pages} {064431} (\bibinfo {year} {2022})}\BibitemShut {NoStop}%
\bibitem [{\citenamefont {Chen}\ \emph {et~al.}(2020)\citenamefont {Chen},
  \citenamefont {Gaudet}, \citenamefont {Dasgupta}, \citenamefont {Marcus},
  \citenamefont {Lin}, \citenamefont {Chen}, \citenamefont {Tomita},
  \citenamefont {Ikhlas}, \citenamefont {Zhao}, \citenamefont {Chen},
  \citenamefont {Stone}, \citenamefont {Tchernyshyov}, \citenamefont
  {Nakatsuji},\ and\ \citenamefont {Broholm}}]{Chen2020}%
  \BibitemOpen
  \bibfield  {author} {\bibinfo {author} {\bibfnamefont {Y.}~\bibnamefont
  {Chen}}, \bibinfo {author} {\bibfnamefont {J.}~\bibnamefont {Gaudet}},
  \bibinfo {author} {\bibfnamefont {S.}~\bibnamefont {Dasgupta}}, \bibinfo
  {author} {\bibfnamefont {G.~G.}\ \bibnamefont {Marcus}}, \bibinfo {author}
  {\bibfnamefont {J.}~\bibnamefont {Lin}}, \bibinfo {author} {\bibfnamefont
  {T.}~\bibnamefont {Chen}}, \bibinfo {author} {\bibfnamefont {T.}~\bibnamefont
  {Tomita}}, \bibinfo {author} {\bibfnamefont {M.}~\bibnamefont {Ikhlas}},
  \bibinfo {author} {\bibfnamefont {Y.}~\bibnamefont {Zhao}}, \bibinfo {author}
  {\bibfnamefont {W.~C.}\ \bibnamefont {Chen}}, \bibinfo {author}
  {\bibfnamefont {M.~B.}\ \bibnamefont {Stone}}, \bibinfo {author}
  {\bibfnamefont {O.}~\bibnamefont {Tchernyshyov}}, \bibinfo {author}
  {\bibfnamefont {S.}~\bibnamefont {Nakatsuji}},\ and\ \bibinfo {author}
  {\bibfnamefont {C.}~\bibnamefont {Broholm}},\ }\bibfield  {title} {\bibinfo
  {title} {Antichiral spin order, its soft modes, and their hybridization with
  phonons in the topological semimetal {Mn}$_3${Ge}},\ }\href
  {https://doi.org/10.1103/PhysRevB.102.054403} {\bibfield  {journal} {\bibinfo
   {journal} {Phys. Rev. B}\ }\textbf {\bibinfo {volume} {102}},\ \bibinfo
  {pages} {054403} (\bibinfo {year} {2020})}\BibitemShut {NoStop}%
\bibitem [{\citenamefont {Chaudhary}\ \emph {et~al.}(2022)\citenamefont
  {Chaudhary}, \citenamefont {Burkov},\ and\ \citenamefont
  {Heinonen}}]{Chaudhary2022}%
  \BibitemOpen
  \bibfield  {author} {\bibinfo {author} {\bibfnamefont {G.}~\bibnamefont
  {Chaudhary}}, \bibinfo {author} {\bibfnamefont {A.~A.}\ \bibnamefont
  {Burkov}},\ and\ \bibinfo {author} {\bibfnamefont {O.~G.}\ \bibnamefont
  {Heinonen}},\ }\bibfield  {title} {\bibinfo {title} {Magnetism and
  magnetotransport in the kagome antiferromagnet {Mn}$_3${Ge}},\ }\href
  {https://doi.org/10.1103/PhysRevB.105.085108} {\bibfield  {journal} {\bibinfo
   {journal} {Phys. Rev. B}\ }\textbf {\bibinfo {volume} {105}},\ \bibinfo
  {pages} {085108} (\bibinfo {year} {2022})}\BibitemShut {NoStop}%
\bibitem [{\citenamefont {Alekhin}\ \emph {et~al.}(2017)\citenamefont
  {Alekhin}, \citenamefont {Razdolski}, \citenamefont {Ilin}, \citenamefont
  {Meyburg}, \citenamefont {Diesing}, \citenamefont {Roddatis}, \citenamefont
  {Rungger}, \citenamefont {Stamenova}, \citenamefont {Sanvito}, \citenamefont
  {Bovensiepen},\ and\ \citenamefont {Melnikov}}]{Alekhin2017}%
  \BibitemOpen
  \bibfield  {author} {\bibinfo {author} {\bibfnamefont {A.}~\bibnamefont
  {Alekhin}}, \bibinfo {author} {\bibfnamefont {I.}~\bibnamefont {Razdolski}},
  \bibinfo {author} {\bibfnamefont {N.}~\bibnamefont {Ilin}}, \bibinfo {author}
  {\bibfnamefont {J.~P.}\ \bibnamefont {Meyburg}}, \bibinfo {author}
  {\bibfnamefont {D.}~\bibnamefont {Diesing}}, \bibinfo {author} {\bibfnamefont
  {V.}~\bibnamefont {Roddatis}}, \bibinfo {author} {\bibfnamefont
  {I.}~\bibnamefont {Rungger}}, \bibinfo {author} {\bibfnamefont
  {M.}~\bibnamefont {Stamenova}}, \bibinfo {author} {\bibfnamefont
  {S.}~\bibnamefont {Sanvito}}, \bibinfo {author} {\bibfnamefont
  {U.}~\bibnamefont {Bovensiepen}},\ and\ \bibinfo {author} {\bibfnamefont
  {A.}~\bibnamefont {Melnikov}},\ }\bibfield  {title} {\bibinfo {title}
  {Femtosecond spin current pulses generated by the nonthermal spin-dependent
  {Seebeck} effect and interacting with ferromagnets in spin valves},\ }\href
  {https://doi.org/10.1103/PhysRevLett.119.017202} {\bibfield  {journal}
  {\bibinfo  {journal} {Phys. Rev. Lett.}\ }\textbf {\bibinfo {volume} {119}},\
  \bibinfo {pages} {017202} (\bibinfo {year} {2017})}\BibitemShut {NoStop}%
\bibitem [{\citenamefont {Ghosh}\ \emph {et~al.}(2012)\citenamefont {Ghosh},
  \citenamefont {Auffret}, \citenamefont {Ebels},\ and\ \citenamefont
  {Bailey}}]{Ghosh2012}%
  \BibitemOpen
  \bibfield  {author} {\bibinfo {author} {\bibfnamefont {A.}~\bibnamefont
  {Ghosh}}, \bibinfo {author} {\bibfnamefont {S.}~\bibnamefont {Auffret}},
  \bibinfo {author} {\bibfnamefont {U.}~\bibnamefont {Ebels}},\ and\ \bibinfo
  {author} {\bibfnamefont {W.~E.}\ \bibnamefont {Bailey}},\ }\bibfield  {title}
  {\bibinfo {title} {Penetration depth of transverse spin current in ultrathin
  ferromagnets},\ }\href {https://doi.org/10.1103/PhysRevLett.109.127202}
  {\bibfield  {journal} {\bibinfo  {journal} {Phys. Rev. Lett.}\ }\textbf
  {\bibinfo {volume} {109}},\ \bibinfo {pages} {127202} (\bibinfo {year}
  {2012})}\BibitemShut {NoStop}%
\bibitem [{\citenamefont {Gomonay}\ and\ \citenamefont
  {Loktev}(2015)}]{Gomonay2015}%
  \BibitemOpen
  \bibfield  {author} {\bibinfo {author} {\bibfnamefont {O.~V.}\ \bibnamefont
  {Gomonay}}\ and\ \bibinfo {author} {\bibfnamefont {V.~M.}\ \bibnamefont
  {Loktev}},\ }\bibfield  {title} {\bibinfo {title} {{Using generalized
  Landau-Lifshitz equations to describe the dynamics of multi-sublattice
  antiferromagnets induced by spin-polarized current}},\ }\href
  {https://doi.org/10.1063/1.4931648} {\bibfield  {journal} {\bibinfo
  {journal} {Low Temp. Phys.}\ }\textbf {\bibinfo {volume} {41}},\ \bibinfo
  {pages} {698} (\bibinfo {year} {2015})}\BibitemShut {NoStop}%
\bibitem [{\citenamefont {Goli}\ and\ \citenamefont
  {Manchon}(2021)}]{Goli2021}%
  \BibitemOpen
  \bibfield  {author} {\bibinfo {author} {\bibfnamefont {V.~M. L. D.~P.}\
  \bibnamefont {Goli}}\ and\ \bibinfo {author} {\bibfnamefont {A.}~\bibnamefont
  {Manchon}},\ }\bibfield  {title} {\bibinfo {title} {Crossover from diffusive
  to superfluid transport in frustrated magnets},\ }\href
  {https://doi.org/10.1103/PhysRevB.103.104425} {\bibfield  {journal} {\bibinfo
   {journal} {Phys. Rev. B}\ }\textbf {\bibinfo {volume} {103}},\ \bibinfo
  {pages} {104425} (\bibinfo {year} {2021})}\BibitemShut {NoStop}%
\bibitem [{\citenamefont {He}\ and\ \citenamefont {Liu}(2024)}]{He2024}%
  \BibitemOpen
  \bibfield  {author} {\bibinfo {author} {\bibfnamefont {Z.}~\bibnamefont
  {He}}\ and\ \bibinfo {author} {\bibfnamefont {L.}~\bibnamefont {Liu}},\
  }\bibfield  {title} {\bibinfo {title} {Magnetic dynamics of strained
  non-collinear antiferromagnet},\ }\href {https://doi.org/10.1063/5.0192467}
  {\bibfield  {journal} {\bibinfo  {journal} {J. Appl. Phys.}\ }\textbf
  {\bibinfo {volume} {135}},\ \bibinfo {pages} {093902} (\bibinfo {year}
  {2024})}\BibitemShut {NoStop}%
\bibitem [{\citenamefont {Yoon}\ \emph {et~al.}(2023)\citenamefont {Yoon},
  \citenamefont {Zhang}, \citenamefont {Chou}, \citenamefont {Takeuchi},
  \citenamefont {Uchimura}, \citenamefont {Hou}, \citenamefont {Han},
  \citenamefont {Kanai}, \citenamefont {Ohno}, \citenamefont {Fukami},\ and\
  \citenamefont {Liu}}]{Yoon2023}%
  \BibitemOpen
  \bibfield  {author} {\bibinfo {author} {\bibfnamefont {J.-Y.}\ \bibnamefont
  {Yoon}}, \bibinfo {author} {\bibfnamefont {P.}~\bibnamefont {Zhang}},
  \bibinfo {author} {\bibfnamefont {C.-T.}\ \bibnamefont {Chou}}, \bibinfo
  {author} {\bibfnamefont {Y.}~\bibnamefont {Takeuchi}}, \bibinfo {author}
  {\bibfnamefont {T.}~\bibnamefont {Uchimura}}, \bibinfo {author}
  {\bibfnamefont {J.~T.}\ \bibnamefont {Hou}}, \bibinfo {author} {\bibfnamefont
  {J.}~\bibnamefont {Han}}, \bibinfo {author} {\bibfnamefont {S.}~\bibnamefont
  {Kanai}}, \bibinfo {author} {\bibfnamefont {H.}~\bibnamefont {Ohno}},
  \bibinfo {author} {\bibfnamefont {S.}~\bibnamefont {Fukami}},\ and\ \bibinfo
  {author} {\bibfnamefont {L.}~\bibnamefont {Liu}},\ }\bibfield  {title}
  {\bibinfo {title} {Handedness anomaly in a non-collinear antiferromagnet
  under spin–orbit torque},\ }\href
  {https://doi.org/10.1038/s41563-023-01620-2} {\bibfield  {journal} {\bibinfo
  {journal} {Nat. Mater.}\ }\textbf {\bibinfo {volume} {22}},\ \bibinfo {pages}
  {1106} (\bibinfo {year} {2023})}\BibitemShut {NoStop}%
\bibitem [{\citenamefont {Xu}\ \emph {et~al.}(2024)\citenamefont {Xu},
  \citenamefont {Zhang}, \citenamefont {Qiao}, \citenamefont {Liang},
  \citenamefont {Shi},\ and\ \citenamefont {Zhu}}]{Xu2024}%
  \BibitemOpen
  \bibfield  {author} {\bibinfo {author} {\bibfnamefont {Z.}~\bibnamefont
  {Xu}}, \bibinfo {author} {\bibfnamefont {X.}~\bibnamefont {Zhang}}, \bibinfo
  {author} {\bibfnamefont {Y.}~\bibnamefont {Qiao}}, \bibinfo {author}
  {\bibfnamefont {G.}~\bibnamefont {Liang}}, \bibinfo {author} {\bibfnamefont
  {S.}~\bibnamefont {Shi}},\ and\ \bibinfo {author} {\bibfnamefont
  {Z.}~\bibnamefont {Zhu}},\ }\bibfield  {title} {\bibinfo {title}
  {Deterministic spin-orbit torque switching including the interplay between
  spin polarization and kagome plane in {Mn}$_3${Sn}},\ }\href
  {https://doi.org/10.1103/PhysRevB.109.134433} {\bibfield  {journal} {\bibinfo
   {journal} {Phys. Rev. B}\ }\textbf {\bibinfo {volume} {109}},\ \bibinfo
  {pages} {134433} (\bibinfo {year} {2024})}\BibitemShut {NoStop}%
\end{thebibliography}%
\end{document}